%% file: founded.tex
\documentclass[11pt]{article}
\usepackage{fullpage}

\newcommand{\full}[1]{#1}
\newcommand{\myconf}[1]{}

\newcommand{\lfcs}[1]{}
\newcommand{\notlfcs}[1]{#1}
\newcommand{\lfcssub}[1]{} %

\newcommand{\popl}[1]{}
\newcommand{\notpopl}[1]{#1}

\newcommand{\tlp}[1]{}
\newcommand{\nottlp}[1]{#1}

\newcommand{\appEx}{Section~\ref{sec-examp}}

\newcommand{\minsp}[1]{#1}

\usepackage{enumitem} %

\usepackage{url}
\usepackage{xspace} 
\usepackage{alltt}
\usepackage[T1]{fontenc}
\usepackage{amsmath,amssymb}

\usepackage{rotating} %

\newcommand{\Hex}[1]{\hspace{#1ex}}
\newcommand{\Vex}[1]{\vspace{#1ex}}

\newenvironment{code}{\begin{alltt}\small}{\end{alltt}}
\newcommand\co[1]{\mbox{\tt\small #1}} %
\newcommand\m[1]{\mbox{$#1$}} %
\newcommand\bigO[1]{\m{O(#1)}} %
\newcommand{\defn}[1]{\textit{#1}} %
\newcommand{\term}[1]{\textit{#1}} %

\def\mathify#1{\m{#1}}
\newcommand\IF{\mathify{\leftarrow}\xspace}
\newcommand\AND{\mathify{\land}\xspace}
\newcommand\NOT{\mathify{\neg}\xspace}
\newcommand\OR{\mathify{\lor}\xspace}

\newcommand\SOME{\mathify{\exists}\xspace}
\newcommand\EACH{\mathify{\forall}\xspace}

\newcommand\T{\mathify{T}\xspace}
\newcommand\F{\mathify{F}\xspace}
\newcommand\UD{\mathify{U}\xspace}
\newcommand\NEQ{\mathify{\neq}\xspace}

\newcommand\NA{--\xspace}

\newcommand\uk[1]{\underline{#1}}
\newcommand\fa[1]{$\overline{\mbox{#1}}$}
\newcommand\cert{certain\xspace}
\newcommand\uncert{uncertain\xspace}

\nottlp{\newcommand{\mysec}[1]{\section{#1}}}
\tlp{}

\nottlp{
\notpopl{\newcommand{\myparag}[1]{\paragraph{#1.}}}
\popl{}
}
\tlp{}

\newenvironment{example}
 {\smallskip\noindent\textbf{\textit{Example}}.}{\hfill $\blacksquare$}

\newcommand{\notes}[1]{} %

\newcommand{\anony}[2]{#1}

\popl{} %

\tlp{}

\newcommand\fund{
    This work was supported in part by NSF under grants 
    CCF-1414078, %
    CNS-1421893, %
    IIS-1447549, %
    CCF-1248184, %
    CCF-0964196, %
    and CCF-0613913, %
    and ONR under grants
    N000142012751,
    N000141512208, %
    and N000140910651. %
}

\begin{document}
\tlp{}

\tlp{}

\tlp{}
\nottlp{
\title{Founded Semantics and Constraint Semantics of 
  Logic Rules\thanks{\fund}
}%
}%
\author{
\anony{
  Yanhong A. Liu \Hex{10} Scott D. Stoller\\
  Computer Science Department, 
  Stony Brook University\\ %
  {\{liu,stoller\}@cs.stonybrook.edu}
}{}
} %
\lfcs{}
\popl{} %

\notpopl{
\date{March 25, 2020}
\maketitle
}
\popl{}

\begin{abstract}
  Logic rules and inference are fundamental in computer science and have
  been studied extensively.  However, prior semantics of logic languages
  can have subtle implications and can disagree significantly, on even 
  very simple programs, 
  including in attempting to solve the well-known Russell's paradox.\lfcssub{}
  These semantics are often non-intuitive and hard-to-understand when
  unrestricted negation is used in recursion.

  This paper describes a simple new semantics for logic rules, {\em founded
    semantics}, and its straightforward extension to another simple new
  semantics, {\em constraint semantics}, that unify the core of different
  prior semantics.  The new semantics support unrestricted negation, as
  well as unrestricted existential and universal quantifications.
  They are uniquely expressive and intuitive by allowing assumptions about
  the predicates, rules, and reasoning to be specified explicitly,
  as simple and precise binary choices.
  They are completely declarative %
  and relate cleanly to prior semantics.
  In addition, founded semantics can be computed in linear time in the
  size of the ground program.

\lfcs{}
\end{abstract}
\vspace{-2ex}

\popl{} %

\nottlp{\Vex{2}\noindent {\bf Keywords:}}
\tlp{}
  Datalog, recursion,
  unrestricted negation, %
  existential and universal quantifications,
  fixed-point semantics,
  constraints,
  well-founded semantics,
  stable model semantics. 
\tlp{}
\tlp{}

\popl{}

\mysec{Introduction}
\label{sec-intro}

Logic rules and inference are fundamental in computer science, especially
for solving complex modeling, reasoning, and analysis problems in critical
areas, such as decision support, program analysis, verification, and security, 
and for knowledge representation and reasoning in general.

The semantics of logic rules and their efficient computations have been a
subject of significant study, especially for complex rules that involve
recursion and unrestricted negation and quantifications.
Many different semantics and computation methods have been proposed,
e.g., see surveys~\cite{apt1994negation,fitting2002fixpoint}.
Unfortunately, different semantics can disagree significantly, 
on even very simple programs.
They are often non-intuitive and hard-to-understand when 
unrestricted negation is used in recursion.
Even those used in many Prolog-based systems and Answer Set
Programming systems---negation as failure~\cite{clark78},
well-founded semantics (WFS)~\cite{van+91well}, and stable model semantics
(SMS)~\cite{GelLif88}---can have subtle implications and differ significantly.
Is it possible to create a simple semantics that also unifies these
different semantics?

In practice, different semantics may be useful under different assumptions
about the facts, rules, and reasoning used.  For example, an application
may have complete information about some predicates, i.e., sets and
relations, but not other predicates.  Capturing such situations is
important for increasingly larger and more complex applications.  Any
semantics that is based on a single set of assumptions for all predicates
cannot best model such applications.  How can a semantics be created to
support all different assumptions and still be simple and easy to use?

\lfcssub{}

This paper describes a simple new semantics for logic rules, {\em founded
  semantics}, and its straightforward extension to another simple new
semantics, {\em constraint semantics}. %
\begin{itemize}

\item The new semantics support unrestricted negation (both stratified
  and non-stratified), as well as unrestricted combinations of
  existential and universal quantifications.

\item They allow each predicate to be specified explicitly as \cert
  (each assertion of the predicate has one of two values: true, false)
  or \uncert (has one of three values: true, false, undefined), and as
  complete (all rules %
  defining the predicate are given) or not.

\item Completion rules are added for predicates that are complete, as
  explicit rules for inferring the negation of those predicates using
  the %
  negation of the hypotheses of the given rules.

\item Founded semantics
  infers all true and false values that
  are founded, i.e., rooted in the given true or false values and
  exactly following the rules, %
  and it completes \cert predicates with false values and completes
  \uncert predicates with undefined values.

\item Constraint semantics\notes{} extends founded semantics by
  allowing undefined values to take all combinations of true and false
  values that satisfy the constraints imposed by the rules.

\notes{}
\end{itemize}

Founded semantics and constraint semantics unify the core of previous
semantics and have three main advantages:
\begin{enumerate}

\item They are expressive and intuitive, by allowing assumptions about
  predicates and rules to be specified explicitly, by including the
  choice of \uncert predicates to support common-sense reasoning with
  ignorance, and by adding explicit completion rules to define the
  negation of predicates.

\item They are completely declarative. %
  Founded
  semantics takes the given rules %
  and completion rules %
  as recursive definitions of the predicates and their negation, and
  is simply the least fixed point of the recursive functions.
  Constraint semantics takes the given rules and completion rules as
  constraints, and is simply the set of all solutions that are
  consistent with founded semantics.

\item They relate cleanly to prior semantics,
  including stratified
  semantics~\cite{apt88}, %
  first-order logic, Fitting semantics (also called Kripke-Kleene
  semantics)~\cite{fitting85}, supported
  models~\cite{apt88}, %
  as well as WFS and SMS, by precisely capturing\lfcssub{} corresponding assumptions about the predicates and
  rules.

\end{enumerate}
Additionally, founded semantics can be computed %
in linear time in the size of the ground
program, %
as opposed to quadratic time for WFS.

Finally, founded semantics and constraint semantics can be extended to
allow \uncert, complete predicates to be specified as closed---making
an assertion of the predicate false
if inferring it to be true (respectively false) 
requires assuming itself to be true (respectively false)---and thus match
WFS and SMS, respectively.

\full{%
The rest of the paper is organized as follows.
Section~\ref{sec-motiv} gives informal motivation for the new semantics.
Section~\ref{sec-lang} describes the rule language.
Sections~\ref{sec-sem}, \ref{sec-prop}, and \ref{sec-cmp} present the
formal definition of the new semantics, properties of the semantics,
and comparison with other semantics, respectively.
Section~\ref{sec-small} compares with other semantics for well-known small examples and more.
Section~\ref{sec-ext} describes linear-time computation, closed
predicate assumption, and other extensions.
Section~\ref{sec-examp} discusses additional well-known examples, showing
that we obtain the desired semantics
for all of them.
Section~\ref{sec-related} discusses related work and concludes.
Appendix~\ref{sec-proofs} contains proofs of all theorems.
} %

This paper is an extended and revised version of Liu and
Stoller~\cite{LiuSto18Founded-LFCS}.  The main changes are 
new Theorem~\ref{thm:equiv-decls} in Section~\ref{sec-prop} to help simplify some results,
new Section~\ref{sec-small} on small examples that was in an appendix, 
new Section~\ref{sec-examp} that describes additional well-known examples, and 
new Appendix~\ref{sec-proofs} that contains complete proofs of all nineteen theorems.

\mysec{Motivation for founded semantics and constraint semantics}
\label{sec-motiv}

Founded semantics and constraint semantics are designed to be
intuitive and expressive.
For rules with no negation or with restricted negation, which have
universally accepted semantics, the new semantics are consistent with
the accepted semantics.  For rules with unrestricted negation, which so far
lack a universally accepted semantics,
the new semantics unify the core of prior semantics with two basic
principles:
\begin{enumerate}

\item Assumptions about certain and uncertain predicates, with true
  (\T) and false (\F) values, or possibly undefined (\UD) values, and
  about whether the rules defining each predicate are complete must be
  made explicit.

\item Any easy-to-understand %
  semantics must be consistent with one where
  everything %
  inferred that has a unique \T or \F value %
  is rooted in the given \T or \F values and following the rules.

\end{enumerate}
This section gives informal explanations.

\myparag{Rules with no negation}
Consider a set of rules with no negation in the hypotheses, %
e.g., a rule can be ``\co{q(x) if p(x)}'' but not ``\co{q(x) if not p(x)}'' for
predicates \co{p} and \co{q} and variable \co{x}.
The meaning of the rules, given a set of facts, e.g., a fact \co{p(a)} for
constant \co{a},
is the set of all facts that are given or can be inferred by applying the
rules to the facts, e.g., \{\co{p(a),q(a)}\} using the example rule and fact
given.
In particular, 
\begin{enumerate}

\item Everything is either \T or \F, i.e., \T as given or inferred facts,
  or \F as otherwise.  So one can just explicitly express what are \T, and
  the rest are \F.

\item Everything inferred must be founded, %
  i.e., rooted in the given facts and following the rules.  So anything
  that always depends on itself, e.g., \co{p(a)}, given only the rule 
  ``\co{p(x) if p(x)}'',
  is not \T.

\end{enumerate}
In technical terms, the semantics is \term{2-valued}, %
and the set of all facts, i.e., true assertions, is the \term{minimum
  model}, equal to the \term{least fixed point} of applying the rules
starting from the given facts.

\myparag{Rules with restricted negation}
Consider rules with negation in the hypotheses, but %
with each negation only
on a predicate all of whose facts can be inferred without using rules
that contain negation of that predicate, %
e.g., one can have ``\co{q(x) if not p(x)}'' %
but not ``\co{p(x) if not p(x)}''.
The meaning of the rules is as for rules with no negation except that a
rule with negation is applied only after all facts of the negated
predicates have been inferred.  In other words,
\begin{enumerate}

\item[] The true assertions of any predicate do not depend on the negation
  of that predicate.
  So a negation could be just a test after all facts of the negated
  predicate are inferred.
  The rest remains the same as for rules with no negation.

\end{enumerate}
In technical terms, this %
is \term{stratified negation};
the semantics is still 2-valued, the minimum model, and the set of all true
assertions is %
the least fixed point of applying the rules in order of the \term{strata}.

\myparag{Rules with unrestricted negation}
Consider rules with unrestricted negation in the hypotheses, where a
predicate may cyclically depend on its own negation, e.g., ``\co{p(x) if not
  p(x)}''.
Now 
the value of a negated assertion needs to be established 
before all facts of the negated predicate have been inferred.
In particular,
\begin{itemize}

\item[] %
  There may not be a unique \T or \F value for each assertion.  For
  example, given only rule ``\co{p(x) if not p(x)}'', \co{p(a)}
  cannot be \T because inferring it following the rule would require itself
  be \F,
  and it cannot be \F because it would lead to itself being \T
  following the rule.
  That is, there may not be a 2-valued model.

\end{itemize}
In technical terms, the negation may be \term{non-stratified}.  There are
two best
solutions to this that generalize a unique 2-valued model: a unique
3-valued model and a set of 2-valued models, as in %
well-founded semantics (WFS) and stable model semantics (SMS),
respectively.

In a unique 3-valued model, when a unique \T or \F value cannot be
established for an assertion, a third value, \defn{undefined} (\UD), is
used.  For example, given only rule
``\co{p(x) if not p(x)}'', \co{p(a)} is \UD,
in both WFS %
and founded semantics.
\begin{itemize}

\item With the semantics being 3-valued, when one cannot infer that an
  assertion is \T, %
  one should be able to express whether it is \F or \UD when there is a
  choice. %
  For example, given only rule ``\co{p(x) if p(x)}'', \co{p(a)} is not \T,
  so \co{p(a)} may in general be \F or \UD.
  
\item WFS %
  requires that such an assertion be \F, even though common sense generally
  says that it is \UD.  WFS %
  attempts to be the same as in the case of 2-valued semantics, even though
  one is now in a 3-valued situation.

\item %
  Founded semantics supports both, %
  allowing one to choose
  explicitly when there is a choice.
  Founded semantics is more expressive by supporting the choice.
  It is also more
  intuitive %
  by supporting the common-sense choice for expressing ignorance.

\end{itemize}

For a set of 2-valued models, similar considerations motivate our
constraint semantics.
In particular, given only rule ``\co{p(x) if not p(x)}'', the semantics is
the empty set, i.e., there is no model, in both SMS %
and constraint semantics, because no model can contain \co{p(a)} or \co{not
  p(a)}, for any \co{a}, because \co{p(a)} cannot be \T or \F as discussed
above.
However, given only rule ``\co{p(x) if p(x)}'', 
SMS requires that \co{p(a)} be \F in all models, whereas constraint semantics
allows the choice of \co{p(a)} being \F in all models or being \T in some
models and \F in other models.

\myparag{Certain or uncertain}
Founded semantics and constraint semantics first allow a predicate to
be declared \term{\cert}
(i.e., each assertion of the predicate has one of two values: \T, \F)
or \term{\uncert}
(i.e., each assertion of the predicate has one of three values: \T, \F,
\UD) when there is a choice.
If a predicate is defined (as conclusions of rules) with use of
non-stratified negation, then it must be declared \uncert, because it
might not have a unique 2-valued model.
Otherwise, it may be declared \cert or \uncert.
\begin{itemize}

\item For a \cert predicate, everything \T must be given or inferred
  by following the rules, and the rest are \F, 
  in both founded semantics and constraint semantics.

\item For an \uncert predicate, everything \T or \F must be given or
  inferred, and the rest are \UD in founded semantics.  Constraint
  semantics then extends everything \UD to be combinations of \T and \F
  that satisfy all the rules and facts as constraints.

\end{itemize}\vspace{-1ex}

\myparag{Complete or not}
Founded semantics and constraint semantics then allow an \uncert
predicate %
to be declared \term{complete},
i.e., all rules with that predicate in the conclusion are given.
\begin{itemize}

\item If a predicate is complete, then completion rules are added to define
  the %
  negation of the predicate explicitly using the %
  negation of the hypotheses of all given rules and facts of that
  predicates.

\item Completion rules, if any, and given rules are used together to infer
  everything \T and \F.  The rest are \UD in founded semantics, and are
  combinations of \T and \F in constraint semantics as described above.

\end{itemize}\vspace{-1ex}

\myparag{Closed or not}
Finally, founded semantics and constraint semantics can be extended to
allow an \uncert, complete predicate to be declared \term{closed},
i.e., an assertion of the predicate is made \F, called
\term{self-false},
if
inferring it to be \T (respectively \F) 
requires assuming itself to be \T (respectively \F).
\begin{itemize}

\item Determining self-false assertions is similar to determining unfounded
  sets in WFS.
  Repeatedly computing founded semantics and self-false assertions until a
  least fixed point is reached yields WFS.

\item Among combinations of \T and \F values for assertions with \UD values
  in WFS, removing each combination that has self-false assertions that are
  not already \F in that combination yields SMS.
\end{itemize}\vspace{-1ex}

\myparag{Correspondence to prior semantics, more on motivation}
Table~\ref{tab-summary} summarizes corresponding declarations 
that capture different assumptions under prior semantics; formal
definitions and proofs for these and for additional relationships appear in
the following sections.
Note that founded semantics and constraint semantics allow additional combinations of
declarations besides those in the table, because different predicates can have different declarations; founded semantics and constraint semantics are exponentially more expressive in this sense.

\begin{table*}%
{
  \centering
  \small
\begin{\nottlp{tabular}\tlp{}}{@{}c@{}\notlfcs{|@{}c@{}}||@{}c@{}|@{\,}c@{\,}|@{\,}c@{\,}|@{\,}c@{\,}||@{}c@{}}
  Prior Semantics \notlfcs{& Kinds of Rules} & New
  & Certain? & Complete? & Closed? & Theorem\tlp{}\\
  \hline\hline

  Stratified
  \notlfcs{& 
    \begin{\nottlp{tabular}\tlp{}}{@{~}c@{~}}
    no non-stratified\\  negation \end{\nottlp{tabular}\tlp{}}}
  & \begin{\nottlp{tabular}\tlp{}}{@{}c@{}}
    Founded\tlp{}\\\hline
    \tlp{}
    \,Constraint\, \end{\nottlp{tabular}\tlp{}}
  & yes
  & \begin{\nottlp{tabular}\tlp{}}{@{}c@{}}
    (implied\\ yes) \end{\nottlp{tabular}\tlp{}}
  & \begin{\nottlp{tabular}\tlp{}}{@{}c@{}}
    (implied\\ yes) \end{\nottlp{tabular}\tlp{}}
  & \ref{thm:comparison-stratified}\tlp{}\\
  \hline\hline

  First-Order Logic
  \notlfcs{& any}
  & Constraint
  & no
  & no
  & \begin{\nottlp{tabular}\tlp{}}{@{}c@{}}
    (implied\\ no) \end{\nottlp{tabular}\tlp{}}
  & \ref{thm:comparison-logic}\tlp{}\\
  \hline\hline

  \begin{\nottlp{tabular}\tlp{}}{@{}c@{}}
    Fitting (Kripke-Kleene)\,\tlp{}\\\hline\tlp{}
    Supported \end{\nottlp{tabular}\tlp{}}
  \notlfcs{& any}
  & \begin{\nottlp{tabular}\tlp{}}{@{}c@{}}
    Founded\tlp{}\\\hline\tlp{}
    \,Constraint\, \end{\nottlp{tabular}\tlp{}}
  & %
    ~no\m{^*}
  & yes
  & no
  & \begin{\nottlp{tabular}\tlp{}}{@{}c@{}}
    \ref{thm:comparison-fitting-equal}\tlp{}\\\hline
    \tlp{}
    \hspace{3.7ex}\ref{thm:comparison-supported-equal}\hspace{3.7ex}
    \end{\nottlp{tabular}\tlp{}}\tlp{}\\
  \hline\hline

  \begin{\nottlp{tabular}\tlp{}}{@{\,}c@{\,}}
    WFS\tlp{}\\\hline\tlp{}
    \hspace{9.7ex}SMS\hspace{9.7ex}
  \end{\nottlp{tabular}\tlp{}}
  \notlfcs{& any}
  & \begin{\nottlp{tabular}\tlp{}}{@{}c@{}}
    Founded\tlp{}\\\hline\tlp{}
    \,Constraint\, \end{\nottlp{tabular}\tlp{}}
  & ~~\,no\m{^{**}}
  & yes
  & yes
  & \begin{\nottlp{tabular}\tlp{}}{@{}c@{}}
    \ref{thm:self-false-WFS}\tlp{}\\\hline
    \tlp{}
    \hspace{3.7ex}\ref{thm:self-false-SMS}\hspace{3.7ex}
    \end{\nottlp{tabular}\tlp{}}\tlp{}\\
  \hline

\notes{}%

\end{\nottlp{tabular}\tlp{}}
\minsp{\vspace{-.75ex}}

} %
\caption{Correspondence between prior semantics and the new semantics, 
  with declarations for all predicates, 
  capturing different assumptions under prior semantics. 
  \m{^*} Some predicates can also be declared certain, 
  per Theorem~\ref{thm:equiv-decls}.
  \m{^{**}} Any predicate can also be declared certain if allowed, 
  per Theorem~\ref{thm:uncertain-certain}.
  \lfcs{}}
\label{tab-summary}
\end{table*}

Some observations from Table~\ref{tab-summary} may help one better understand founded
semantics and constraint semantics.
\begin{itemize}

\item The\lfcssub{} 4 wide rows cover all combinations of allowed
  declarations (if all predicates have the same declarations).

\item Wide row 1 is a special case of wide row 4, because being \cert
  implies being complete and closed.  So one could prefer to use only
  the latter two choices and omit the first choice.  However, being
  \cert is uniquely important, both for conceptual simplicity and
  practical efficiency:

  (1)~ It covers the vast class of database applications that %
  do not use non-stratified
  negation, for which stratified semantics is universally accepted.
  It does not need to be understood by explicitly combining the latter two
  more sophisticated notions.

  (2)~ It allows founded semantics 
  to match WFS for all example programs we found in the
  literature, %
  with predicates being \cert when possible and complete otherwise,
  but without the last, most sophisticated notion of being closed; and
  the semantics can be computed in linear time.

\item Wide rows 2, 3, and 4 allow the assumptions about predicates that are
  \uncert, not complete, or not closed 
  to be made explicitly.

\notes{}%

\end{itemize}

In a sense, WFS %
uses \F for both false and some kinds of ignorance (no knowledge of
something must mean it is \F), uses \T for both true and some kinds of
ignorance inferred %
through negation of \F, and uses \UD for conflict, remaining kinds of
ignorance from \T and \F, and imprecision;
SMS %
resolves the ignorance in \UD, but not the ignorance in \F and \T.  In
contrast,
\begin{itemize}

\item founded semantics %
  uses \T only for true, \F only for false, and \UD for conflict, ignorance,
  and imprecision;

\item constraint semantics %
  further differentiates among conflict, ignorance, and
  imprecision---\linebreak[1]cor\-responding to there being no model, multiple models, and
  a unique model, respectively, consistent with founded semantics.

\end{itemize}

After all, any easy-to-understand semantics must be consistent with the \T
and \F\notes{} assertions that can be inferred by exactly
following the rules and completion rules starting from the given
facts. %
\begin{itemize}

\item Founded semantics is the maximum set of such \T and \F assertions, as
  a least fixed point of the given rules and completion rules if any, plus
  \UD\lfcssub{} for the remaining assertions.

\item Constraint semantics is the set of combinations of all \T and \F
  assertions %
  that are consistent with founded semantics and satisfy the rules as
  constraints.

\end{itemize}
Founded semantics without closed predicates can be computed easily and
efficiently, as a least fixed point, contrasting with an alternating fixed
point or iterated fixed point for computing WFS.

\mysec{Language}
\label{sec-lang}

We first consider %
Datalog with unrestricted negation in hypotheses.
We extend it in Section~\ref{sec-ext} to allow unrestricted
combinations of existential and universal quantifications 
and other features.

\myparag{Datalog with unrestricted negation}
A \defn{program} in the core language is a finite set of rules of the
following form, where any \m{P_i} may be preceded with \NOT, and any
\m{P_i} and \m{Q} over all rules may be declared \cert or \uncert, and
declared complete or not: %
\minsp{\vspace{-.75ex}}
\begin{equation}
  \label{eqn-rule}
  Q(X_1,\ldots,X_a) ~\IF~ 
  P_1(X_{11},\ldots,X_{1a_1}) ~\AND~ \cdots ~\AND~ P_h(X_{h1},\ldots,X_{ha_h})
\end{equation}
Symbols \IF, \AND, and \NOT indicate backward implication, conjunction, and
negation, respectively; \m{h} is a natural number and is possibly 0, each \m{P_i}
(respectively \m{Q}) is a predicate\notes{} of finite number
\m{a_i} (respectively \m{a}) of arguments, each \m{X_{ij}} and \m{X_k} is
either a constant or a variable, and each variable in the arguments of
\m{Q} must also be in the arguments of some \m{P_i}.

If \m{h = 0}, there is no \m{P_i} or \m{X_{ij}}, and each \m{X_k} must be a
constant, in which case \m{Q(X_1, \ldots, X_a)} is called a \defn{fact}.
For the rest of the paper, ``rule'' refers only to the case where \m{h \geq
  1}, in which case each \m{P_i(X_{i1}, \ldots, X_{ia_i})} or \m{\NOT
  P_i(X_{i1}, \ldots, X_{ia_i})} is called a \defn{hypothesis} of the rule,
and \m{Q(X_1, \ldots, X_a)} is called the \defn{conclusion} of the rule.
The set of hypotheses of the rule is called the \defn{body} of the rule.

A predicate declared \cert %
means that each assertion of the predicate has a unique true (\T) or
false (\F) value.  A predicate declared \uncert means that each
assertion of the predicate has a unique true, false, or undefined (\UD)
value.
A predicate declared complete means that
all rules with that predicate in the conclusion are given in the
program. 

A predicate in the conclusion of a rule is said to be \defn{defined}
using the predicates or their negation in the hypotheses of the rule,
and this defined-ness relation is transitive.
\begin{itemize}

\item A predicate must be declared \uncert
  if it is defined transitively using its own negation, or is defined
  using an \uncert predicate; otherwise, it may be declared \cert or
  \uncert and is by default \cert.

\item A predicate may be declared complete or not only if it is
  \uncert, %
and it is by default complete.

\end{itemize}
In examples with no explicit specification of declarations, default
declarations are used.

Rules of form~(\ref{eqn-rule}) without negation are captured exactly by
Datalog~\cite{Ceri:Gottlob:Tanca:90,AbiHulVia95}, a database query
language based on the logic programming paradigm.  Recursion in
Datalog allows queries not expressible in relational algebra or
relational calculus. %
Negation allows more sophisticated logic to
be expressed directly.  However, unrestricted negation in recursion
has been the main challenge in defining the semantics of such a
language,
e.g.,~\cite{apt1994negation,fitting2002fixpoint}, %
including whether the semantics should be 2-valued or 3-valued.

\begin{example}
  We use %
  \co{win}, the win-not-win game, as a running example, with default
  declarations: \co{move} is \cert, and \co{win} is \uncert and
  complete.
  A move from position \co{x} to position \co{y} is represented by a fact
  \co{move(x,y)}.  The following rule captures the win-not-win game: a
  position \co{x} is winning if there is a move from \co{x} to some position
  \co{y} and \co{y} is not winning.
  Arguments \co{x} and \co{y} are variables.%
\begin{code}
    win(x) \IF move(x,y) \AND \NOT win(y)
\end{code}%
Note that the declarations for predicates \co{move} and \co{win} are
different.  Other choices of declarations can lead to different results,
e.g., see the first example in Section~\ref{sec-examp}.
\end{example}

\myparag{Notations}
In arguments of predicates, %
we use letter sequences for variables, and use numbers and quoted strings
for constants.

In presenting the semantics, in particular the completion rules, we use
equality and the notations below for existential and universal
quantifications, respectively, in the hypotheses of rules, and use negation
in the conclusions.
\minsp{\vspace{-.75ex}}
\begin{equation}
  \label{eq:quant}
\begin{\nottlp{tabular}\tlp{}}[c]{@{}ll@{}}
  $\SOME~ X_1, \ldots, X_n~\co{|}~ Y$ & ~~existential quantification\\
  $\EACH~ X_1, \ldots, X_n~\co{|}~ Y$ & ~~universal quantification
\end{\nottlp{tabular}\tlp{}}
\end{equation}
The quantifications return \T iff for some or all, respectively,
combinations of values of $X_1,\ldots,X_n$, the value of Boolean expression
$Y$ is \T.
The domain of each quantified variable is the set of all constants in the
program.

\mysec{Formal definition of founded semantics and constraint semantics}
\label{sec-sem}

\input semantics

\input{smallexamp.tex}

\mysec{Computational complexity and extensions}
\label{sec-ext}

\tlp{}
\myparag{Computing founded semantics and constraint semantics}\notpopl{~\\}
\begin{mytheorem}
  Computing founded semantics is linear time in the size of the ground
  program. %
\end{mytheorem}

\begin{myproof}{}
  First ground all given rules, using any grounding.  Then add completion
  rules, if any, %
  by adding an inverse rule for each group of the grounded given rules that
  have the same conclusion, yielding ground completion rules of the same
  asymptotic size as the grounded given rules.

  Now compute the least fixed point for each SCC
  of the resulting ground rules using a previous
  method~\cite{LiuSto09Rules-TOPLAS}.  To do so, first introduce a new
  intermediate predicate and rule for each conjunction and disjunction in
  the rules, yielding a new set of rules of the same asymptotic size.  In
  computing the least fixed point, each resulting rule incurs at most
  one %
  rule firing %
  because there are no variables in the rule, and each firing takes
  worst-case \bigO{1} time.  Thus, the total time is worst-case linear in
  the size of all ground rules and therefore in the size of the grounded
  given rules.
\end{myproof}

The size of the ground program is polynomial in the size \m{n} of input
data, i.e., the given facts, because each variable in each rule can be
instantiated at most \bigO{n} times (because the domain size is at most
\m{n}), and there is a fixed number of variables in each rule, and a fixed
size of the given rules.  Precisely, the size of the ground program is in
the worst case \bigO{n^k\times r}, where \m{k} is the maximum number of
variables in a rule, and \m{r} is the size of the given rules.

Computing constraint semantics may take exponential time in the size %
of the input data, because 
in the worst case, all assertions of all predicates may have \UD values in
founded semantics, and there is an exponential number of combinations of \T
and \F values of all assertions, where each combination may be checked for
whether it satisfies the constraints imposed by all rules.

These complexity analyses also apply to the extensions below except that
computing founded semantics with closed predicates may take quadratic time
in the size of the ground program, because of repeated computation of
founded semantics and self-false assertions.

\selfFalseSemantics
\notes{}%

Note, however, that founded semantics for default declarations (\cert
when possible and complete otherwise) allows the number of repetitions
for computing self-false atoms to be greatly reduced, even to zero,
compared with WFS that does
repeated computation of unfounded sets.

In all examples we have found in the literature, and all natural
examples we have been able to think of, founded semantics for
default declarations, without closed predicate assumption, infers the
same result as WFS.
However, while founded semantics computes a single least fixed point
without the outer repetition and is worst-case linear time, WFS computes an
alternating fixed point or iterated fixed point and is worst-case
quadratic.
In fact, we have not found any natural example showing that an actual
quadratic-time alternating or iterated fixed-point for computing WFS
is needed.\footnote{Even a contrived example that demonstrates the
  worst-case quadratic-time computation of WFS has been challenging to
  find.  For example, the quadratic-time example
  in~\cite{zukowski2001thesis} turns out to be linear in XSB; after
  significant effort between us and Warren, we found a much more
  sophisticated example that appears to take quadratic time, but a
  remaining bug in XSB makes the correctness of its computation
  unclear.}

\notes{}

\notes{}%

\myparag{Unrestricted\full{
existential and universal} quantifications in hypotheses}
We extend the language to allow unrestricted combinations of existential
and universal quantifications as well as negation, conjunction, and
disjunction in hypotheses.  The domain of each quantified variable is the
set of all constants in the program.  \notes{}

\begin{example}
  For the \co{win} example, the following two rules may be given instead:%
\begin{code}
    win(x) \IF \SOME y | move(x,y) \AND lose(y)
    lose(x) \IF \EACH y | \NOT move(x,y) \OR win(y)
\end{code}\vspace{-3ex}
\end{example}

The semantics in Section~\ref{sec-sem} is easily extended to accommodate
this extension: these constructs simply need to be interpreted, using their
3-valued logic semantics~\cite{fitting85}, when defining one-step
derivability.
Theorems~\ref{thm:consistent}--\full{\ref{thm:pred-merge}}\myconf{}
hold for this extended language.  The other semantics discussed in Section~\ref{sec-cmp} are not defined for this extension, thus we do not have theorems relating to them.

\myparag{Negation in facts and conclusions}
We extend the language to allow negation in given facts and in
conclusions of given rules; such facts and rules are said to be {\em
  negative}.  The Yale shooting example in \appEx\ is a simple example.  

The definition of founded semantics applies directly to this extension, because it already introduces and handles negative rules, and it already infers and handles negative facts.  Note that $\comb$ combines only positive facts and positive rules to form combined rules; negative facts and negative rules are copied unchanged into the completed program.

With this extension, a program and hence its founded model may be
inconsistent; for example, a program could contain or imply \co{p} and
$\NOT \co{p}$.  Thus, Theorem \ref{thm:consistent} does not hold for
such programs.
When the founded model is inconsistent, the inconsistent literals in it can
easily be reported\full{ to the user}.  When the founded model is
consistent, the definition of constraint semantics applies directly, and
Theorems
\ref{thm:model}--\full{\ref{thm:pred-merge}}\myconf{}
hold.  The other semantics discussed in Section~\ref{sec-cmp} are not
defined for this extended language, so we do not have theorems relating to
them.

\notes{}

\input{addexamp.tex}

\mysec{Related work and conclusion}
\label{sec-related}

There is a large literature on logic language semantics
\lfcs{}
\notlfcs{and implementations.} %
Several overview
articles~\cite{apt1994negation,prz94well,ramUll95survey,fitting2002fixpoint}
give a good sense of the challenges when there is unrestricted negation.
We discuss major prior semantics here\notlfcs{; major implementations are as
discussed in Section~\ref{sec-intro}}.

\nottlp{Clark~\cite{clark78}}\tlp{} 
describes completion of logic programs to give a semantics
for negation as failure.
Numerous others,
e.g.,~\cite{lloyd84making,sato84transformational,Jaffar+86some,chan88constructive,foo88deduced,stuckey91constructive},
describe similar additions.
\nottlp{Fitting~\cite{fitting85}}\tlp{}
presents a semantics, called Fitting semantics or
Kripke-Kleene semantics, that aims to give a least 3-valued model.
\nottlp{Apt et al.~\cite{apt88}}\tlp{}
defines supported model semantics, which is a set
of 2-valued models; the models correspond to extensions of the Fitting model.
\nottlp{Apt et al.~\cite{apt88}}\tlp{}
introduces stratified semantics.
WFS~\cite{van+91well} %
also gives a 3-valued model but aims to maximize false values.
SMS~\cite{GelLif88} also gives a set of 2-valued models and aims to
maximize false values.
Other formalisms and semantics include partial stable models, also called
stationary models~\cite{prz94well}, and
FO(ID), %
for first-order logic with inductive
definitions%
~\cite{denecker2008logic}.\notes{}
There are also many studies that relate different semantics,
e.g.,~\cite{dung1992relations,lin2004assat}.

Our founded semantics, which extends to constraint semantics, is unique in
that it allows predicates
to be specified as \cert or \uncert,
as complete or not, %
and as closed or not.
These choices clearly and explicitly capture the different assumptions
one can have about the predicates, rules, and reasoning, 
and capture them as simple and precise binary choices, unlike the
well-known closed-world assumption and open-world assumption,
and allow different combinations of assumptions to co-exist naturally.
These choices make founded and constraint semantics
more expressive and intuitive.
Instead of using many separate semantics,
one just need to make the assumptions explicit; the
same underlying logic is used for inference.
In this way, founded semantics and constraint semantics unify different
semantics.

In addition, founded semantics and constraint semantics are completely
declarative,
as a least fixed point and as constraint satisfaction, respectively. %
Our default declarations without closed predicates lead to the same
semantics as WFS and SMS %
for all natural examples we have found.
Additionally, founded semantics without closed predicates can be
computed in linear time in the size of the ground
program, %
as opposed to quadratic time for WFS.

Liu and Stoller~\cite{LiuSto20LogicalConstraint-LFCS} creates a unified
language for design and analysis, DA logic, based on founded semantics and
constraint semantics, to support the power and ease of programming with
different intended semantics.
It provides meta-constraints to specify different assumptions, supports the
use of uncertain information in the form of either undefined values or
possible combinations of values, and, for composability, introduces
knowledge units that can be instantiated by any new predicates, including
predicates with additional arguments.

There are many directions for future study, including additional
relationships with prior semantics, further extensions,
efficient implementations, %
and applications.

\newcommand{\ack}{
\minsp{\vspace{1ex}}
\myparag{Acknowledgment}
We thank David S.\ Warren, Michael Kifer, Anil Nerode, 
Tuncay Tekle, Molham Aref, %
Marc Denecker, %
Cordell Green, Goyal Gupta, %
John Hooker,
Bob Kowalski, 
Fangzhen Lin,
Zuoquan Lin,
Alberto Pettorossi, Maurizio Proietti, 
Carlo Zaniolo,
Neng-Fa Zhou, and many others for helpful
comments and discussions
on logic languages, semantics, and efficient computations.
}
\ack

\tlp{}

\def\bibdir{../../bib}                  %
{
\let\OLDthebibliography\thebibliography
\renewcommand\thebibliography[1]{
  \OLDthebibliography{#1}
  \setlength{\parskip}{0pt}
  \setlength{\itemsep}{4.2pt plus 0.3ex}
}
\bibliography{\bibdir/strings,\bibdir/liu,\bibdir/IC,\bibdir/PT,\bibdir/Lang,\bibdir/Algo,\bibdir/DB,\bibdir/Sys,\bibdir/misc,\bibdir/crossref}
\nottlp{\notpopl{\notlfcs{\bibliographystyle{alpha}}}}
\lfcs{}
\tlp{} %
}

\appendix

\lfcs{}

\input{appendix-proofs}

\end{document}

%% file: semantics.tex
\newcommand{\proofs}[1]{#1}
\newcommand{\noproofs}[1]{}



\newcounter{thmcounter}
\newenvironment{mytheorem}{
\refstepcounter{thmcounter}
\noindent\textbf{\textit{Theorem \thethmcounter}}.}{\hfill}

\newenvironment{myproof}[1]{{\bf Proof#1. }}{\hfill$\blacksquare$}
\newcommand{\union}{\cup}
\newcommand{\intersect}{\cap}
\newcommand{\pgm}{\pi}
\newcommand{\comb}{{\it Combine}}
\newcommand{\addinv}{{\it AddInv}}
\newcommand{\cmpl}{{\it Cmpl}}
\newcommand{\lfp}{{\it LFP}}
\newcommand{\lfpscc}{{\it LFPbySCC}}
\newcommand{\addneg}{{\it AddNeg}}
\newcommand{\ccmpl}{{\it CCmpl}}
\newcommand{\fitting}{{\it Fitting}}
\newcommand{\wf}{{\it WFS}}
\newcommand{\founded}{{\it Founded}}
\newcommand{\constraint}{{\it Constraint}}
\newcommand{\stratified}{{\it Stratified}}
\newcommand{\supported}{{\it Supported}}
\newcommand{\stable}{{\it SMS}}
\newcommand{\foid}{{\it FOID}}
\newcommand{\dg}{{\it DG}}
\newcommand{\nameneg}{{\it NameNeg}}
\newcommand{\unnameneg}{{\it UnNameNeg}}
\newcommand{\proj}[2]{{\it Proj}(#1,#2)}
\newcommand{\merge}{{\it Merge}}
\newcommand{\mpa}{{\it MergeAtom}}
\newcommand{\selffalse}{{\it SelfFalse}}
\newcommand{\foundedcl}{{\it FoundedClosed}}
\newcommand{\constraintcl}{{\it ConstraintClosed}}
\newcommand{\atom}{{\it Atom}}
\newcommand{\ufs}{U_0}
\newcommand{\UA}{{\it UA}}


\tlp{\vspace{2ex}}
\myparag{Atoms, literals, consistency, and projection}
Let $\pgm$ be a program.
A predicate is {\em intensional} in $\pgm$ if it appears in the conclusion of at least one rule; otherwise, it is {\em extensional}.  

An {\em atom} of $\pgm$ is a formula formed by applying a predicate symbol in $\pgm$ to constants in $\pgm$.  A {\em literal} of $\pgm$ is an atom of $\pgm$ or the negation of an atom of $\pgm$.  These are called {\em positive literals} and {\em negative literals}, respectively.  
The literals $p$ and $\neg p$ are {\em complements} of each other.  A set of literals is {\em consistent} if it does not contain a literal and its complement.

The {\em projection} of a program $\pgm$ onto a set $S$ of predicates, denoted $\proj{\pgm}{S}$, contains all facts of $\pgm$ whose predicates are in $S$ and all rules of $\pgm$ whose conclusions contain predicates in $S$. 

\myparag{Interpretations, ground instances, models, and derivability}
An {\em interpretation} of $\pgm$ is a consistent set of literals of $\pgm$.  Interpretations are generally 3-valued: a literal $p$ is {\em true} (\T) in interpretation $I$ if it is in $I$, is {\em false} (\F) in $I$ if its complement is in $I$, and is {\em undefined} (\UD) in $I$ if neither it nor its complement is in $I$.  An interpretation of $\pgm$ is {\em 2-valued} if it contains, for each atom $A$ of $\pgm$, either $A$ or its complement.  An interpretation $I$ is {\em 2-valued for predicate $P$} if, for each atom $A$ for $P$, $I$ contains $A$ or its complement.  Interpretations are ordered by set inclusion $\subseteq$.

A {\em ground instance} of a rule $R$ is any rule that can be obtained from $R$ by expanding universal quantifications into conjunctions over all constants in the domain, and then instantiating the remaining variables with constants.  
For example, $\co{q(a)}~\IF~\co{p(a)}~\AND~\co{r(b)}$ is a ground instance of $\co{q(x)}~\IF~\co{p(x)}~\AND~\SOME~\co{y}~|~\co{r(y)}$.  

An interpretation is a {\em model} of a program if it contains all facts in the program and satisfies all rules of the program, regarded as formulas in 3-valued logic \cite{fitting85}, i.e., for each ground instance of each rule, if the body is true, then so is the conclusion.  

The {\em one-step derivability} operator $T_\pgm$ for program $\pgm$ performs one step of inference using rules of $\pgm$, starting from a given interpretation.  Formally, $C\in T_\pgm(I)$ iff $C$ is a fact of $\pgm$ or there is a ground instance $R$ of a rule of $\pgm$ with conclusion $C$ such that each hypothesis of $R$ is true in interpretation~$I$.  


\myparag{Dependency graph}
The {\em dependency graph} $\dg(\pgm)$ of program $\pgm$ is a directed graph with a node for each predicate of $\pgm$, and an edge from $Q$ to $P$ labeled $\mathord{+}$ (respectively, $\mathord{-}$) if a rule whose conclusion contains $Q$ has a positive (respectively, negative) hypothesis that contains $P$.  If the node for predicate $P$ is in a cycle containing only positive edges, then $P$ has {\it circular positive dependency} in $\pgm$; if it is in a cycle containing a negative edge, then $P$ has {\em circular negative dependency} in $\pgm$.




\subsection{Founded semantics}

Intuitively, the {\em founded model} of a program $\pgm$, denoted $\founded(\pgm)$, is the least set of literals that are given as facts or can be inferred by repeated use of the rules. 
We define $\founded(\pgm) = \unnameneg(\lfpscc(\nameneg(\cmpl(\pgm))))$, where 
functions $\cmpl$, $\nameneg$, $\lfpscc$, and $\unnameneg$ are defined as follows.

\myparag{Completion}
The completion function, $\cmpl(\pgm)$, returns the {\it completed
  program} of $\pgm$.  Formally, $\cmpl(\pgm)=\addinv(\comb(\pgm))$, where
$\comb$ and $\addinv$ are defined as follows.

The function $\comb(\pgm)$ returns the program obtained from $\pgm$ by replacing the facts and rules defining each \uncert complete predicate $Q$ with a single {\em combined rule} for $Q$, defined as follows.  Transform the facts and rules defining $Q$ so they all have the same conclusion $Q(V_1, \ldots, V_a)$, where $V_1,\ldots,V_a$ are fresh variables (i.e., not occurring in the given rules defining $Q$), by replacing each fact or rule $Q(X_1, \ldots, X_a) ~\IF~ H_1 \land \cdots \land H_h$ with $Q(V_1,\ldots,V_a) ~\IF~ (\SOME~Y_1,\ldots,Y_k ~|~\tlp{\linebreak} V_1=X_1 \land \cdots \land V_a=X_a \land H_1 \land \cdots \land H_h)$, where  $Y_1,\ldots,Y_k$ are all variables occurring in the given fact or rule.  Combine the resulting rules for $Q$ into a single rule defining $Q$ whose body is the disjunction of the bodies of those rules.  This combined rule for $Q$ is logically equivalent to the original facts and rules for $Q$.  Similar completion rules are used in Clark completion \cite{clark78} and Fitting semantics \cite{fitting85}.

\begin{example} 
  For the \co{win} example with default declarations, the rule for \co{win} becomes the following.  For readability, we renamed variables to transform the equality conjuncts into tautologies and then eliminated them.
\begin{code}
    win(x) \IF \SOME y | (move(x,y) \AND \NOT win(y))
\end{code}\vspace{-3ex}
\end{example}



The function $\addinv(\pgm)$ returns the program obtained from $\pgm$ by adding, for each  \uncert complete predicate $Q$, a {\em completion rule} that derives negative literals for $Q$.  The completion rule for $Q$ is obtained from the inverse of the combined rule defining $Q$ (recall that the inverse of $C~\IF~B$ is $\neg C~\IF~\neg B$), by putting the body of the rule in negation normal form, i.e., using laws of predicate logic to move negation inwards and eliminate double negations, so that negation is applied only to atoms.




\begin{example}
For the \co{win} example with default declarations, the added rule is
\begin{code}
    \NOT win(x) \IF \EACH y | (\NOT move(x,y) \OR win(y))
\end{code}\vspace{-3ex}
\end{example}

\myparag{Least fixed point}
%
%
%
The least fixed point is preceded and followed by functions that introduce and remove, respectively, new predicates representing the negations of the original predicates.


The function $\nameneg(\pgm)$ returns the program obtained from $\pgm$ by
replacing each negative literal $\neg P(X_1, \ldots, X_a)$ with
$\co{n.}P(X_1, \ldots, X_a)$, where the new predicate $\co{n.}P$ represents
the negation of predicate $P$.

\begin{example}
For the \co{win} example with default declarations, this yields:
\begin{code}
    win(x) \IF \SOME y | (move(x,y) \AND n.win(y))
    n.win(x) \IF \EACH y | (n.move(x,y) \OR win(y))
\end{code}\vspace{-3ex}
\end{example}

The function $\lfpscc(\pgm)$ uses a least fixed point to infer facts for each strongly connected component (SCC) in the dependency graph of $\pgm$, as follows.  Let $S_1,\ldots,S_n$ be a list of the SCCs in dependency order, so earlier SCCs do not depend on later ones; it is easy to show that any linearization of the dependency order leads to the same result for $\lfpscc$.  
For convenience, we overload $S$ to also denote the set of predicates in the SCC.

Define $\lfpscc(\pgm) = I_n$, where $I_0$ is the empty set and $I_i =
\addneg(\lfp(\myconf{\linebreak[1]}T_{I_{i-1} \union \proj{\pgm}{S_i}}), S_i)$ for $i \in
1..n$.  $\lfp$ is the least fixed point operator.  The least fixed point
is well-defined, because the one-step derivability function $T_{I_{i-1}
  \union \proj{\pgm}{S_i}}$ is monotonic, because the program $\pgm$ does
not contain negation.  The function $\addneg(I, S)$ returns the interpretation obtained from interpretation $I$ by adding {\em completion facts} for \cert predicates in $S$ to $I$; specifically, for each \cert predicate $P$ in $S$, for each combination of values $v_1,\ldots,v_a$ of arguments of $P$, if $I$ does not contain $P(v_1,\ldots,v_a)$, then add $\co{n.}P(v_1,\ldots,v_a)$.


\begin{example}
  For the \co{win} example with default declarations, the least fixed point calculation
  \begin{enumerate}

  \item infers \co{n.win(x)} for any \co{x} that does not have
    \co{move(x,y)} for any \co{y}, i.e., has no move to anywhere;

  \item infers \co{win(x)} for any \co{x} that has \co{move(x,y)} for
    some \co{y} and \co{n.win(y)} has been inferred;

  \item infers more \co{n.win(x)} for any \co{x} such that any \co{y}
    having \co{move(x,y)} has \co{win(y)};

  \item repeatedly does 2 and 3 above until a fixed point is reached.
  \end{enumerate}\vspace{-3ex}









\end{example}

The function $\unnameneg(I)$ returns the interpretation obtained from interpretation $I$ by replacing each atom $\co{n.}P(X_1, \ldots, X_a)$ with $\neg P(X_1, \ldots, X_a)$.

\begin{example}
For the \co{win} example with default declarations, positions \co{x} for which \co{win(x)} is \T, \F, and \UD, respectively, in the founded model correspond exactly to the well-known winning, losing, and draw positions, respectively.  In particular,
  \begin{enumerate}

  \item a losing position is one that either does not have a move to
    anywhere or has moves only to winning positions; 
  
  \item a winning position is one that has a move to a losing position; and

  \item a draw position is one not satisfying either case above, i.e., it
    is in a cycle of moves that do not have a move to a losing position,
    called a {\em draw cycle}, or is a position that has only sequences of
    moves to positions in draw cycles.

  \end{enumerate}\vspace{-3ex}

\end{example}

\subsection{Constraint semantics}

Constraint semantics is a set of 2-valued models based on founded semantics.  A {\em constraint model} of $\pgm$ is a consistent 2-valued interpretation $M$ such that $M$ is a model of $\cmpl(\pgm)$ and $\founded(\pgm) \subseteq M$.  We define $\constraint(\pgm)$ to be the set of constraint models of $\pgm$.  Constraint models can be computed from $\founded(\pgm)$ by iterating over all assignments of true and false to atoms that are undefined in $\founded(\pgm)$, and checking which of the resulting interpretations satisfy all rules in $\cmpl(\pgm)$.


\begin{example}
  For the \co{win} example with default declarations, draw positions (i.e., positions for which \co{win} is undefined) are in draw cycles, i.e., cycles that do not have a \co{move} to a \co{n.win} position, or are positions that have only a sequence of moves to positions in draw cycles.
  \begin{enumerate}
  
\item If some SCC has draw cycles of only odd lengths, then there is
  no satisfying assignment of \T and \F to \co{win} for positions in
  the SCC, so there are no constraint models of the program.

\item If some SCC has draw cycles of only even lengths, then there are
  two satisfying assignments of \T and \F to \co{win} for positions in
  the SCC, with the truth values alternating between \T and \F around
  each cycle, and with the second truth assignment obtained from the
  first by swapping \T and \F.
  The total number of constraint models of such SCCs is exponential
  in the number of such SCCs.
  \end{enumerate}\vspace{-3ex}
\end{example}

\mysec{Properties of founded semantics and constraint semantics}
\label{sec-prop}

Proofs of theorems appear in\proofs{\nottlp{ Appendix}
  \ref{sec-proofs}}\noproofs{ \appProofs}.

\myparag{Consistency and correctness}
The most important properties are consistency and correctness.

\begin{mytheorem}
  The founded model and constraint models of a program $\pgm$ are consistent.
  \label{thm:consistent}
\end{mytheorem}

\newcommand{\thmConsistentProof}{
\noindent
\begin{myproof}{ of Theorem \ref{thm:consistent}}
  First we show that the founded model is consistent.  A given program cannot contain negative facts or negative conclusions, so all negative literals in $\founded(\pgm)$ are added by the construction.  For a predicate declared \uncert and not complete, no negative literals are added.  For a predicate $P$ declared \uncert and complete, 
consistency follows from the fact that the only rule defining \co{n.}$P$ in $\cmpl(\pgm)$ is the inverse of the only rule defining $P$ in $\cmpl(\pgm)$.
The body of the former rule is the negation of the body $B$ of the latter rule.  Monotonicity of $T_{I_{i-1} \union \proj{\nameneg(\cmpl(\pgm))}{S_i}}$ implies that the value of a ground instance of $B$ cannot change from true to false, or {\it vice versa}, during the fixed point calculation for the SCC $S$ containing $P$.  Using this observation, it is easy to show by induction on the number of iterations of the fixed point calculation for $S$ that an atom for $P$ and its negation cannot both be added to the interpretation.
For a \cert predicate, consistency follows from the fact that $\addneg$ adds only literals whose complement is not in the interpretation.  

Constraint models are consistent by definition.
\end{myproof}}

\begin{mytheorem}
  The founded model of a program $\pgm$ is a model of $\pgm$ and $\cmpl(\pgm)$.  The constraint models of $\pgm$ are 2-valued models of $\pgm$ and $\cmpl(\pgm)$.
  \label{thm:model}
\end{mytheorem}

\newcommand{\thmModelProof}{
\noindent
\begin{myproof}{ of Theorem \ref{thm:model}}
  First we show that $\founded(\pgm)$ is a model of $\cmpl(\pgm)$. $\founded(\pgm)$ contains all facts in $\pgm$, because each fact in $\pgm$ is either merged into a combined rule in $\cmpl(\pgm)$ or copied unchanged into $\cmpl(\pgm)$, and in either case is added to the founded model by the LFP for some SCC.  Consider a rule $C~\IF~B$ in $\cmpl(\pgm)$ with predicate $Q$ in the conclusion $C$. Note that $C$ may be a positive or negative literal.  If the body $B$ becomes true before or in the LFP for the SCC $S$ containing $Q$, then the corresponding disjunct in the combined rule defining $Q$ becomes true before or in that LFP, so the conclusion $C$ is added to the interpretation by that LFP, so the rule is satisfied.  It remains to show that $B$ could not become true after that LFP.  $B$ cannot become true during processing of a subsequent SCC, because SCCs are processed in dependency order, so subsequent SCCs do not contain predicates in $B$.  We prove by contradiction that $B$ cannot become true in $\addneg$ for $S$, i.e., we suppose $B$ becomes true in $\addneg$ for $S$ and show a contradiction.  $\addneg$ for $S$ adds only negative literals for \cert predicates  in $S$, so $B$ must contain such a literal, say $\neg P(\ldots)$.  $P$ and $Q$ are in the same SCC $S$, so $P$ must be defined, directly or indirectly, in terms of $Q$.  Since $P$ is \cert and is defined in terms of $Q$, $Q$ must be \cert.  Since $Q$ and $P$ are defined in the same SCC $S$, and $Q$ depends negatively on $P$, $Q$ has a circular negative dependency, so $Q$ must be \uncert, a contradiction.


Constraint models are 2-valued models of $\cmpl(\pgm)$ by definition.

Any model of $\cmpl(\pgm)$ is also a model of $\pgm$, because $\pgm$ is logically equivalent to the subset of $\cmpl(\pgm)$ obtained by removing the completion rules added by $\addinv$.
\end{myproof}}



\myparag{Equivalent declarations}

When a predicate can be declared certain, other declarations sometimes have the same effect.

\begin{mytheorem}
  Let $P$ be a predicate that can be declared certain in program $\pgm$.  Let $S$ be the set containing $P$ and the predicates on which $P$ depends.  If each predicate in $S$ does not have positive circular dependency in $\pgm$, then the founded semantics of $\pgm$ is the same if predicates in $S$ are certain or are uncertain and complete. 
\label{thm:equiv-decls}
\end{mytheorem}

\newcommand{\thmEquivDeclsProof}{
\noindent
\begin{myproof}{ of Theorem \ref{thm:equiv-decls}}
Since $P$ can be declared certain, predicates in $S$ do not have negative circular dependency, so they can be declared certain.  Let $\pgm$ be the program with all predicates in $S$ declared certain.  Let $M = \founded(\pgm)$.  Let $\pgm'$ be the variant of $\pgm$ with all predicates in $S$ declared uncertain and complete.  Let $M' = \founded(\pgm')$.  Let $\prec$ be the restriction to $S$ of the dependency relation between predicates in $\pgm$.  The premise about absence of positive circular dependency implies $(S, \prec)$ is acyclic.  We prove by induction on $(S, \prec)$ that, for each predicate $Q$ in $S$, $M$ and $M'$ contain the same literals for $Q$.

Recall that the interpretation of $Q$ in $M$ is calculated as follows.  In the LFP for the SCC containing $Q$, atoms for $Q$ are inferred (i.e., added to $M$) using the facts and rules for $Q$ in $\pgm$; after that LFP, every remaining atom for $Q$ is set to false by $\addneg$ (i.e., the atom's negation is added to $M$).  Note that $M$ is 2-valued for $Q$.

Consider the calculation of the interpretation of $Q$ in $M'$.  In the LFP for the SCC containing $Q$, atoms for $Q$ are inferred using the combined rule for $Q$, and atoms for $\co{n.}P$ are inferred using the completion rule for $Q$.  The combined rule for $Q$ is logically equivalent to the facts and rules for $Q$ in $\pgm$, so $M$ and $M'$ contain the same positive literals for $Q$.  Next, we show that $M'$ (like $M$) is 2-valued for $Q$; this and the fact that they contain the same positive literals for $Q$ imply that they also contain the same negative literals for $Q$, and hence they contain the same literals for $Q$, as desired.  By the induction hypothesis, $M'$ contains the same literals as $M$ for all predicates (if any) on which $Q$ depends, hence $M'$ is 2-valued for those predicates.  This implies that, for every ground instance $R$ of the combined rule for $Q$, either the body of $R$ evaluates to true in $M'$, or the body of the corresponding ground instance $R_{\it comp}$ of the completion rule for $Q$ (i.e., $R_{\it comp}$ is the inverse of $R$) evaluates to true.  This implies that $M'$ is 2-valued for $Q$.
\end{myproof}}

\myparag{Same SCC, same certainty}

All predicates in an SCC have the same certainty.

\begin{mytheorem}
  For every program, for every SCC $S$ in its dependence graph, all predicates in $S$ are \cert, or all of them are \uncert.
  \label{thm:scc-certainty}
\end{mytheorem}

\newcommand{\thmSCCcertaintyProof}{
\noindent
\begin{myproof}{ of Theorem \ref{thm:scc-certainty}}
  It suffices to show that, if some predicate in $S$ is \uncert, then all predicates in $S$ are \uncert.  Suppose $S$ contains an \uncert predicate $P$, and let $Q$ be another predicate in $S$.  $Q$ is defined directly or indirectly in terms of predicate $P$, and $P$ is \uncert, so $Q$ must be \uncert.
\end{myproof}}

\myparag{Higher-order programming}
%
%
%
Higher-order logic programs, in languages such as HiLog, can be encoded as
first-order logic programs by a semantics-preserving transformation that
replaces uses of the original predicates with uses of a single predicate
\co{holds} whose first argument is the name of an original predicate
\cite{CheKW93}.  For example, \co{win(x)} is replaced with
\co{holds(win,x)}.  This transformation merges a set of predicates into a
single predicate, facilitating higher-order programming.  We show that
founded semantics and constraint semantics are preserved by merging of {\em
  compatible} predicates, defined below, if a simple type system is used to
distinguish the constants in the original program from the new constants
representing the original predicates.

We extend the language with a simple type system.  A type denotes a set of constants.  Each predicate has a type signature that specifies the type of each argument.  A program is well-typed if, in each rule or fact, (1) each constant belongs to the type of the argument where the constant occurs, and (2) for each variable, all its occurrences are as arguments with the same type.  In the semantics, the values of predicate arguments are restricted to the appropriate type.


Predicates of program $\pgm$ are {\em compatible} if they are in the same SCC in $\dg(\pgm)$ and have the same arity, same type signature, and (if \uncert) same completeness declaration.  For a set $S$ of compatible predicates of program $\pgm$ with arity $a$ and type signature $T_1,\ldots,T_a$, the {\em predicate-merge transformation} $\merge_S$ transforms $\pgm$ into a program $\merge_S(\pgm)$ in which predicates in $S$ are replaced with a single fresh predicate \co{holds} whose first parameter ranges over $S$, and which has the same completeness declaration as the predicates in $S$.  Each atom $A$ in a rule or fact of $\pgm$ is replaced with $\mpa_S(A)$, where the function $\mpa_S$ on atoms is defined by: $\mpa_S(P(X_1, \ldots, X_a))$ equals \co{holds('$P$', $X_1$, \ldots, $X_a$)} if $P \in S$ and equals $P(X_1, \ldots, X_a)$ otherwise.  We extend $\mpa_S$ pointwise to a function on sets of atoms, used for properties of founded semantics, and a function on sets of sets of atoms, used for properties of constraint semantics.  The predicate-merge transformation introduces $S$ as a new type.  The type signature of \co{holds} is $S, T_1,\ldots, T_a$.



\begin{mytheorem}
  Let $S$ be a set of compatible predicates of program $\pgm$.  Then  $\merge_S(\pgm)$ and $\pgm$ have the same founded semantics, in the sense that $\founded(\merge_S(\pgm)) = \mpa_S(\founded(\pgm))$.  $\merge_S(\pgm)$ and $\pgm$ also have the same constraint semantics, in the sense that 
$\constraint(\merge_S(\pgm))$ ${}= \mpa_S(\constraint(\pgm))$.
  \label{thm:pred-merge}
\end{mytheorem}

\newcommand{\thmPredMergeProof}{
\noindent
\begin{myproof}{ of Theorem \ref{thm:pred-merge}}
  The proof is based on a straightforward correspondence between the constructions of founded semantics of $\pgm$ and $\merge_S(\pgm)$.  

Note that:
\begin{itemize}
\item All predicates in $S$ are \cert, or all of them are \uncert, by Theorem \ref{thm:scc-certainty}.

\item There is a 1-to-1 correspondence between the set of disjuncts in the bodies of the rules for predicates in $S$ in $\cmpl(\pgm)$ and the set of disjuncts in the body of the rule for \co{holds} in $\cmpl(\merge_S(\pgm))$.

\item If predicates in $S$ are \uncert and complete, there is a 1-to-1 correspondence between the set of conjuncts in the bodies of the completion rules for predicates in $S$ in $\cmpl(\pgm)$ and the set of conjuncts in the body of the completion rule for \co{holds} in $\cmpl(\merge_S(\pgm))$.
\end{itemize}

Based on these observations, it is straightforward to show that:
\begin{itemize}
\item For each predicate $P$ not in $S$, each atom $A$ for $P$ or $\co{n}.P$ is derivable in the semantics for $\pgm$ iff $A$ is derivable in the semantics for $\merge_S(\pgm)$.

\item In the LFP for the SCC containing $S$, for each predicate $P$ in $S$, an atom $A$ for $P$ is derivable using a disjunct of the rule for $P$ in $\cmpl(\pgm)$ iff $\mpa_S(A)$ is derivable using the corresponding disjunct of the rule for \co{holds} in $\cmpl(\merge_S(\pgm))$.


\item In the LFP for the SCC containing $S$, for each \uncert complete predicate $P$ in $S$, an atom $A$ for $\co{n}.P$ is derivable using the completion rule for $P$ in $\pgm$ iff $\mpa_S(A)$ is derivable using the corresponding conjuncts in the completion rule for \co{holds} in $\merge_S(\pgm)$ (the other conjuncts in the completion rule for \co{holds} have the form \co{v $\ne$ '$Q$' \OR $\cdots$} and hence are true when considering derivation of atoms of the form \co{n.holds('$P$', $\ldots$))}.

\item In $\addneg$ for the SCC containing $S$, for each \cert predicate $P$  in $S$, an atom $A$ for $\co{n}.P$ is inferred in the semantics for $\pgm$ iff $\mpa_S(A)$ is inferred in the semantics for $\merge_S(\pgm)$.
\end{itemize}
\end{myproof}}

\mysec{Comparison with other semantics}
\label{sec-cmp}
\vspace{1ex}

\tlp{\vspace{2ex}}
\myparag{Stratified semantics}
%
%
\full{ A program $\pgm$ has {\em stratified negation} if it does not contain predicates with circular negative dependencies.  Such a program has a well-known and universally accepted semantics that defines a unique 2-valued model, denoted $\stratified(\pgm)$, as discussed in Section \ref{sec-motiv}.}\myconf{Let $\stratified(\pgm)$ denote the unique 2-valued model of a program with stratified negation, as discussed in Section \ref{sec-motiv}.}


\begin{mytheorem}
  For a program $\pgm$ with stratified negation and in which all predicates are \cert, $\founded(\pgm)=\stratified(\pgm)$.
  \label{thm:comparison-stratified}
\end{mytheorem}

\newcommand{\thmComparisonStratifiedProof}{
\noindent
\begin{myproof}{ of Theorem \ref{thm:comparison-stratified}}
  For \cert predicates, the program completion $\cmpl$ has no effect, and $\lfpscc$ is essentially the same as the definition of stratified semantics, except using SCCs in the dependency graph instead of strata.  The SCCs used in founded semantics subdivide the strata used in stratified semantics; intuitively, this is because predicates are put in different SCCs whenever possible, while predicates are put in different strata only when necessary.  This subdivision of strata does not affect the result of $\lfpscc$, so founded semantics is equivalent to the stratified semantics.
\end{myproof}}

\myparag{First-order logic}
The next theorem relates constraint models with the meaning of a program regarded as a set of formulas in first-order logic; recall that the definition of a model of a program also regards rules as logical formulas.

\begin{mytheorem}
  For a program $\pgm$ in which all predicates are \uncert and not complete, the constraint models of $\pgm$ are exactly the 2-valued models of $\pgm$.
  \label{thm:comparison-logic}
\end{mytheorem}

\newcommand{\thmComparisonLogicProof}{
\noindent
\begin{myproof}{ of Theorem \ref{thm:comparison-logic}}
  Observe that, for a program $\pgm$ satisfying the hypotheses of the theorem, $\cmpl(\pgm)$ is logically equivalent to $\pgm$.
  Every constraint model is a 2-valued model of $\cmpl(\pgm)$ and hence a 2-valued model of $\pgm$.  Consider a 2-valued model $M$ of $\pgm$.  Since $\pgm$ satisfies the hypotheses of the theorem, $\founded(\pgm)$ contains only positive literals, added by the LFPs in $\lfpscc$.  The LFPs add a positive literal to $\founded(\pgm)$ only if that literal is implied by the facts and rules in $\pgm$ and therefore holds in all 2-valued models of $\pgm$.  Therefore, $\founded(\pgm)\subseteq M$.  $M$ satisfies $\pgm$ and hence, by the above observation, also $\cmpl(\pgm)$. Thus, $M$ is a constraint model of $\pgm$.
\end{myproof}}



\myparag{Fitting semantics}
\newcommand{\fittingBackground}{
Fitting \cite{fitting85} defines an interpretation to be a model of a program iff it satisfies a formula that we denote as $\ccmpl(\pgm)$.  This formula is Fitting's 3-valued-logic version of the Clark completion of $\pgm$ \cite{clark78}.  Briefly, $\ccmpl(\pgm) = \ccmpl_D(\pgm) \land \ccmpl_U(\pgm)$, where $\ccmpl_D(\pgm)$ is the conjunction of formulas corresponding to the combined rules introduced by $\comb$ except with $\IF$ replaced with $\cong$ (which is called ``complete equivalence'' and means ``same truth value''), and $\ccmpl_U(\pgm)$ is the conjunction of formulas stating that predicates not used in any fact or the conclusion of any rule are false for all arguments.  The {\em Fitting model} of a program $\pgm$, denoted $\fitting(\pgm)$, is the least model of $\ccmpl(\pgm)$ \cite{fitting85}.
}
\full{\fittingBackground}
\myconf{The {\em Fitting model} of a program $\pgm$, denoted $\fitting(\pgm)$, is the least model of a formula in 3-valued logic \cite{fitting85}; \appOtherSem\ summarizes the definition. }


\begin{mytheorem}
 For a program $\pgm$ in which all predicates are \uncert and complete, $\founded(\pgm)=\fitting(\pgm)$.
  \label{thm:comparison-fitting-equal}
\end{mytheorem}

\newcommand{\thmComparisonFittingEqualProof}{
\noindent
\begin{myproof}{ of Theorem \ref{thm:comparison-fitting-equal}}
  Consider an intensional predicate $P$.  It is straightforward to show that the LFP for the SCC containing $P$ using the combined rule for $P$ in $\cmpl(\pgm)$, of the form $C~\IF~B$, and its inverse, of the form $\neg C~\IF~\neg B$, is equivalent to satisfying the conjunct for $P$ in $\ccmpl_D(\pgm)$, of the form $C \cong B$.  The proof for the forward direction ($\Rightarrow$) of the equivalence is a case analysis on the truth value of the body $B$ in $\founded(\pgm)$: (1) if $B$ is true, then the LFP uses the combined rule $C~\IF~B$ to infer $C$ is true, so $C \cong B$ holds; (2) if $B$ is false, then the LFP uses the inverse rule to infer $C$ is false, so $C \cong B$ holds; (3) if $B$ is undefined, then neither rule applies and $C$ is undefined, so $C \cong B$ holds.  Similarly, the proof for the reverse direction ($\Leftarrow$) is a simple case analysis on the truth values of $B$ and $C$ (which are the same, since $C \cong B$ by assumption).

  Consider an extensional predicate $P$.  Let $S$ be the set of atoms for $P$ in $\pgm$.  It is easy to show that $\founded(\pgm)$ and $\fitting(\pgm)$ contain the atoms in $S$ and contain negative literals for $P$ for all other arguments.
\end{myproof}}

\full{\noindent Theorem \ref{thm:equiv-decls} implies that Theorem \ref{thm:comparison-fitting-equal} also holds if predicates that do not have positive circular dependency and can be declared certain are declared certain instead of uncertain and complete.

Founded semantics for some declarations is less defined than or equal to Fitting semantics, as stated in the following theorem.

\begin{mytheorem}
  (a) For a program $\pgm$ in which all intensional predicates are \uncert and complete, $\founded(\pgm) \subseteq \fitting(\pgm)$.  (b) If, furthermore, some extensional predicate is \uncert, and some positive literal $p$ for some \uncert extensional predicate does not appear in $\pgm$, then $\founded(\pgm) \subset \fitting(\pgm)$.
%
  \label{thm:comparison-fitting-subset}
\end{mytheorem}

\newcommand{\thmComparisonFittingSubsetProof}{
\noindent
\begin{myproof}{ of Theorem \ref{thm:comparison-fitting-subset}}
  (a) This follows from Theorem \ref{thm:comparison-fitting-equal} and the observation that, if $\pgm$ satisfies the premises of Theorem \ref{thm:comparison-fitting-equal}, and $\pgm'$ is obtained from $\pgm$ by changing the declarations of some extensional predicates from \cert to \uncert, then $\founded(\pgm') \subseteq \founded(\pgm)$; intuitively, fewer assumptions are made about \uncert predicates, so $\founded(\pgm')$ contains fewer conclusions.

(b) This follows from part (a) and the observation that $p$ is undefined in $\founded(\pgm)$, and $p$ is false in $\fitting(\pgm)$ (i.e., $\fitting(\pgm)$ contains $\neg p$), so the inclusion relation is strict.
\end{myproof}}
\noindent A simple program $\pgm_6$ for which the inclusion in Theorem \ref{thm:comparison-fitting-subset} is strict, as in part (b) of the theorem, is program 6 in Table \ref{tab-sem}, which has only one rule \co{q \IF p}.  With both predicates being \uncert and complete, $\founded(\pgm_6)=\emptyset$ and $\fitting(\pgm_6)=\{\neg \co{p}, \neg \co{q}\}$.}
  
\full{Founded semantics for default declarations is at least as defined as
  Fitting semantics, as stated in the following theorem.}


\begin{mytheorem}
(a) For a program $\pgm$ in which all predicates have default declarations as \cert or \uncert and complete, $\fitting(\pgm) \subseteq \founded(\pgm)$.
(b) If, furthermore, $\fitting(\pgm)$ is not 2-valued for some \cert intensional predicate $P$, then $\fitting(\pgm) \subset \founded(\pgm)$.
  \label{thm:comparison-fitting-supset}
\end{mytheorem}

\newcommand{\thmComparisonFittingSupsetProof}{
\noindent
\begin{myproof}{ of Theorem \ref{thm:comparison-fitting-supset}}
(a)  This follows from Theorem \ref{thm:comparison-fitting-equal}, the differences between the declarations assumed in Theorem \ref{thm:comparison-fitting-equal} and the default declarations, and the effect of those differences on the founded model.  It is easy to show that the default declarations can be obtained from the declarations assumed in Theorem \ref{thm:comparison-fitting-equal} by changing the declarations of some intensional predicates from \uncert and complete to \cert.  Let $P$ be such a predicate.  This change does not affect the set $S$ of positive literals derived for $P$, because the combined rule for $P$ is equivalent to the original rules and facts for $P$.  This change can only preserve or increase the set of negative literals derived for $P$, because $\addneg$ derives all negative literals for $P$ that can be derived while preserving consistency of the interpretation (in particular, negative literals for all arguments of $P$ not in $S$).

(b) This follows from the proof of part (a) and the observation that the additional premise for part (b) implies there is a literal $p$ for $P$ that is undefined in $\fitting(\pgm)$ and defined (i.e., true or false) in $\founded(\pgm)$ (because $\founded(\pgm)$ is 2-valued for $P$), so the inclusion is strict.
\end{myproof}}
 \noindent A simple program $\pgm_3$ for which the inclusion in Theorem \ref{thm:comparison-fitting-supset} is strict, as in part (b) of the theorem, is program 3 in Table \ref{tab-sem}, which has only one rule \co{q \IF q}. $\fitting(\pgm_3)=\emptyset$ and $\founded(\pgm_3)=\{\NOT \,\co{q}\}$.
  
%

\myparag{Well-founded semantics}
\newcommand{\wellfoundedBackground}{The {\em well-founded model} of a program $\pgm$, denoted $\wf(\pgm)$, is the least fixed point of a monotone operator $W_\pgm$ on interpretations, defined as follows \cite{van+91well}.  A set $\ufs$ of atoms of a program $\pgm$ is an {\em unfounded set} of $\pgm$ with respect to an interpretation $I$ of $\pgm$ iff, for each atom $A$ in $\ufs$, for each ground instance $R$ of a rule of $\pgm$ with conclusion $A$, either (1) some hypothesis of $R$ is false in $I$ or (2) some positive hypothesis of $R$ is in $\ufs$.  Intuitively, the atoms in $\ufs$ can be set to false, because each rule $R$ whose conclusion is in $\ufs$ either has a hypothesis already known to be false or has a hypothesis in $\ufs$ (which will be set to false).  Let $U_\pgm(I)$ be the {\it greatest unfounded set} of program $\pgm$ with respect to interpretation $I$.  For a set $S$ of atoms, let $\neg \cdot S$ denote the set containing the negations of those atoms.  $W_\pgm$ is defined by $W_\pgm(I) = T_\pgm(I) \union \neg \cdot U_\pgm(I)$.  The well-founded model $\wf(\pgm)$ is a model of $\ccmpl(\pgm)$, so $\fitting(\pgm) \subseteq \wf(\pgm)$ for all programs $\pgm$~\cite{van+91well}.
}
\full{\wellfoundedBackground}
\myconf{The {\em well-founded model} of a program $\pgm$, denoted $\wf(\pgm)$, is the least fixed point of a monotone operator $W_\pgm$ on interpretations \cite{van+91well}; \appOtherSem\ summarizes the definition.}


\begin{mytheorem}
  For every program $\pgm$, $\founded(\pgm) \subseteq \wf(\pgm)$.
  \label{thm:comparison-wf}
\end{mytheorem}


\newcommand{\thmComparisonWfProof}{
\noindent
\begin{myproof}{ of Theorem \ref{thm:comparison-wf}}
  We prove an invariant that, at each step during the construction of $\founded(\pgm)$, the current approximation $I$ to $\founded(\pgm)$ satisfies $I \subseteq \wf(\pgm)$.  First we show, using the induction hypothesis, that literals added to $I$ by the LFPs in $\lfpscc$ are in $\wf(\pgm)$.  Consider a literal $p$ added by a combined rule $C~\IF~B$.  This implies $B$ is true in $I$.  By the induction hypothesis, $I \subseteq \wf(\pgm)$, so $B$ is true in $\wf(\pgm)$.  Using the rule in $\pgm$ corresponding to a disjunct in $B$ that is true in $I$, we conclude $p \in T_\pgm(\wf(\pgm))$.  The definition of $\wf(\pgm)$ implies $\wf(\pgm)$ is closed under $T_\pgm$, so $p \in \wf(\pgm)$.  Consider a literal $\neg p$ added by a combined rule $\neg C~\IF~\neg B$.  All of the disjuncts in the negation normal form of $B$ are true in $I$, so the bodies of all rules in $\pgm$ that derive $p$ are false in $I$ and, by the induction hypothesis, are false in $\wf(\pgm)$, so $p \in U_\pgm(\wf(\pgm))$.  The definition of $\wf(\pgm)$ implies $\wf(\pgm)$ is closed under $\neg \cdot U_\pgm$, so $\neg p \in \wf(\pgm)$.

It remains to show that negative literals added to $I$ by $\addneg$ are in $\wf(\pgm)$.  Consider an SCC $S$ in the dependency graph.  Let $N_S$ be the set of atoms whose negations are added to $I$ by $\addneg$ for $S$.  Let $I_S$ denote
the interpretation produced by the LFP for $S$.  Since $U_\pgm$ is monotone, it suffices to show that $N_S$ is an unfounded set for $\pgm$ with respect to $I_S$, i.e., for each atom $A$ in $N_S$, for each ground instance $A~\IF~B$ of a rule of $\pgm$ with conclusion $A$, either (1) some hypothesis in $B$ is false in $I_S$ or (2) some positive hypothesis in $B$ is in $N_S$.  We use a case analysis on the truth value of $B$ in $I_S$.  $B$ cannot be true in $I_S$, because if it were, $A$ would be added to $I_S$ by the LFP and would not be in $N_S$.  If $B$ is false in $I_S$, then case (1) holds.

Suppose $B$ is undefined in $I_S$.  This implies that at least one hypothesis $H$ in $B$ is undefined in $I_S$.
  Let $Q$ be the predicate in $A$, and let $P$ be the predicate in $H$.    $\addneg$ adds literals only for \cert predicates, so $Q$ is \cert.  
$Q$ depends on $P$, so $P$ must be \cert, and $P$ must be in $S$ or a previous SCC.  If $P$ were in a previous SCC, then $I_S$ would be 2-valued for $P$, and $H$ would be $\T$ or $\F$ in $I_S$, a contradiction, so $P$ is in $S$.  Since $P$ is in $S$, and $H$ is undefined in $I_S$, $\addneg$ adds $\neg H$ to $I_S$, i.e., $H$ is in $N_S$.  $Q$ is \cert, so $Q$ does not have circular negative dependency; therefore, since $P$ and $Q$ are both in $S$, $H$ must be a positive hypothesis.  Thus, case (2) holds.
\end{myproof}}
\noindent The inclusion in Theorem \ref{thm:comparison-wf} is strict for program 8 in Table~\ref{tab-sem}, denoted $\pgm_8$, which has only one rule \co{q}~\IF~\NOT\co{q}~\AND~\co{q}.  $\founded(\pgm_8)=\emptyset$ and $\wf(\pgm_8)=\{\neg\co{q}\}$.
 


\myparag{Supported models}
\newcommand{\supportedBackground}{
Supported model semantics of a logic program is a set of 2-valued models.  An interpretation $I$ is a {\it supported model} of $\pgm$ if $I$ is 2-valued and $I$ is a fixed point of the one-step derivability operator $T_\pgm$ \cite{apt88}.  Let $\supported(\pgm)$ denote the set of supported models of $\pgm$.  Supported models, unlike Fitting semantics and WFS, allow atoms to be set to true when they have circular positive dependency and nothing else, like the atom \co{q} in program $\pgm_3$ described above.  }
\full{\supportedBackground}
\myconf{Supported model semantics of a logic program $\pgm$ is a set of 2-valued models \cite{apt88}, denoted $\supported(\pgm)$; \appOtherSem\ summarizes the definition.}

\full{The following three theorems relating constraint semantics with supported model semantics are analogous to the three theorems relating founded semantics with Fitting semantics.}



\begin{mytheorem}
  For a program $\pgm$ in which all predicates are \uncert and complete, $\supported(\pgm)=\constraint(\pgm)$.
\label{thm:comparison-supported-equal}
\end{mytheorem}

\newcommand{\thmComparisonSupportedEqualProof}{
\noindent
\begin{myproof}{ of Theorem \ref{thm:comparison-supported-equal}}
  Let $M\in\supported(\pgm)$.  We show $M\in\constraint(\pgm)$, i.e., $\founded(\pgm) \subseteq M$ and $M$ is a 2-valued model of $\cmpl(\pgm)$.  Theorem 15 of \cite{apt88} shows that an interpretation $I$ is a supported model of $\pgm$ iff $I$ is a 2-valued model of $\ccmpl(\pgm)$ .  Therefore, $M$ is a model of $\ccmpl(\pgm)$.  Theorem \ref{thm:comparison-fitting-equal} implies that $\founded(\pgm)$ is the least model of $\ccmpl(\pgm)$, so $\founded(\pgm) \subseteq M$.  For each predicate $P$ for which $\cmpl(\pgm)$ contains a rule (i.e., each predicate that appears in at least one fact or conclusion in $\pgm$), the conjunction of the combined rule for $P$ and its inverse in $\cmpl(\pgm)$ is logically equivalent for 2-valued models to the equivalence for $P$ in $\ccmpl(\pgm)$; this is a straightforward tautology in 2-valued logic.  Thus, since $M$ is a model of $\ccmpl(\pgm)$, it is also a model of $\cmpl(\pgm)$.
Thus, $M$ is a constraint model of $\pgm$.

Let $M\in\constraint(\pgm)$.  We show that $M\in\supported(\pgm)$.  By definition, $M$ is 2-valued, $M$ satisfies $\cmpl(\pgm)$, and $\founded(\pgm)\subseteq M$.  
By Apt et al.'s theorem cited above, it suffices to show that $M$ is 2-valued and satisfies $\ccmpl(\pgm)$.  We show that the clause in $\ccmpl(\pgm)$ for each predicate $P$ is satisfied, by case analysis on $P$.  

Case 1: $P$ does not appear in any fact or conclusion. 
$\addneg$ makes $P$ false for all arguments in $\founded(\pgm)$ and hence in $M$, so $M$ satisfies the clause for $P$ in $\ccmpl_U(\pgm)$.  Case 2: $P$ appears in some fact or conclusion. 

Case 2: $P$ appears in some fact or conclusion.  $M$ satisfies $\cmpl(\pgm)$, which contains a combined rule of the form $C~\IF~B$ for $P$ and the inverse rule $\NOT C~\IF~\NOT B$.  Since $M$ is 2-valued, the conjunction of those rules is equivalent to $C \cong B$, which is the clause for $P$ in $\ccmpl_D(\pgm)$.
\end{myproof}}



\full{\noindent Theorem \ref{thm:equiv-decls} implies that Theorem \ref{thm:comparison-supported-equal} also holds if predicates that do not have positive circular dependency and can be declared certain are declared certain instead of uncertain and complete.}

\begin{mytheorem}
  For a program $\pgm$ in which all intensional predicates are \uncert and complete, $\supported(\pgm) \subseteq \constraint(\pgm)$.
  \label{thm:comparison-supported-supset}
\end{mytheorem}


\newcommand{\thmComparisonSupportedSupsetProof}{
\noindent
\begin{myproof}{ of Theorem \ref{thm:comparison-supported-supset}}
  This theorem follows from Theorem \ref{thm:comparison-supported-equal}, and the observation that, if $\pgm$ satisfies the premises of Theorem \ref{thm:comparison-supported-equal}, and $\pgm'$ is obtained from $\pgm$ by changing the declarations of some extensional predicates from \cert to \uncert, then $\constraint(\pgm) \subseteq \constraint(\pgm')$.  To prove this, we analyze how the change in declarations affects $\founded(\pgm)$ and $\cmpl(\pgm)$, and then show that the changes to $\founded(\pgm)$ and $\cmpl(\pgm)$ preserve or increase the set of constraint models.  
  
  As shown in the proof of Theorem \ref{thm:comparison-fitting-subset}, the founded model decreases, i.e., $\founded(\pgm') \subseteq \founded(\pgm)$.  This may allow additional constraint models, because more models satisfy the requirement of being a superset of the founded model.  The effect on $\cmpl(\pgm)$ is to add completion rules for the predicates whose declaration changed.  This does not change the set of constraint models, because all constraint models of $\pgm$ satisfy these completion rules.  This conclusion follows from the lemma: For a program $\pgm$ and a \cert predicate $P$ in $\pgm$, every constraint model $M$ of $\pgm$ satisfies the completion rule $R$ for $P$ (even though this rule does not appear in $\cmpl(\pgm)$, because $P$ is \cert).  To see this, first note that $P$ and hence all predicates on which it depends are \cert, so all predicates used in $R$ are \cert, so $\founded(\pgm)$ is 2-valued for those predicates, so $\founded(\pgm)$ and $M$ contain the same literals for those predicates, so $M$ satisfies $R$ iff $\founded(\pgm)$ satisfies $R$.  To see that $\founded(\pgm)$ satisfies $R$, let $C~\IF~B$ denote the combined rule for $P$, so $R$ is $\neg C~\IF~\neg B$, and note that $\founded(\pgm)$ contains a positive literal for $P$ for arguments for which $B$ holds in $\founded(\pgm)$ and contains a negative literal for $P$ for all other arguments, and $B$ is false for all of those other arguments, because $\founded(\pgm)$ is 2-valued for all predicates used in $B$ and hence for $B$.
\end{myproof}}

\noindent The inclusion in Theorem \ref{thm:comparison-supported-supset} is strict for the program $\pgm_6$ described above.  $\supported(\pgm_6) = \{ \{\NOT\,\co{p}, \NOT\,\co{q}\} \}$ and $\constraint(\pgm_6) = \{ \{\co{p}, \co{q}\}, \{\NOT\,\co{p}, \NOT\,\co{q}\} \}$.
     
\begin{mytheorem}
  For a program $\pgm$ in which all predicates have default declarations as \cert or \uncert and complete, $\constraint(\pgm) \subseteq \supported(\pgm)$.
  \label{thm:comparison-supported-subset}
\end{mytheorem}

\newcommand{\thmComparisonSupportedSubsetProof}{
\noindent
\begin{myproof}{ of Theorem \ref{thm:comparison-supported-subset}}
  This theorem follows from Theorem \ref{thm:comparison-supported-equal}, the differences between the declarations assumed in Theorem \ref{thm:comparison-supported-equal} and the default declarations, and the effect of those differences on the constraint models.  We analyze how the differences in declaration affect $\founded(\pgm)$ and $\cmpl(\pgm)$, and then analyze how the changes to $\founded(\pgm)$ and $\cmpl(\pgm)$ affect the set of constraint models.  Specifically, we show that the set of constraint models is preserved or decreases and hence that $\constraint(\pgm) \subseteq \supported(\pgm)$.

  The declarations assumed in Theorem \ref{thm:comparison-supported-equal} are the same as in Theorem \ref{thm:comparison-fitting-equal}.  Recall from the proof of Theorem \ref{thm:comparison-fitting-supset} that the default declarations can be obtained from those declarations by changing the declarations of some intensional predicates from \uncert and complete to \cert.  Let $S$ be the set of predicates whose declarations change.  As discussed in the proof of Theorem \ref{thm:comparison-fitting-supset}, the effect of these declaration changes on $\founded(\pgm)$ is to preserve or increase the set of negative literals for predicates in $S$.  The effect of these declaration changes on $\cmpl(\pgm)$ is to remove completion rules for predicates in $S$.

  Now consider the effects of these changes on the set of constraint models.  Adding negative literals to the founded model has the effect of decreasing the set of constraint models of the program, because constraint models not containing those literals are eliminated, since each constraint model must be a superset of the founded model.  Removing from $\cmpl(\pgm)$ the completion rules for predicates in $S$ does not 
cause any further changes to the set of constraint models, because the interpretation of predicates in $S$ in the constraint models is now completely determined by the requirement that the constraint models are supersets of the founded model, because the founded model is 2-valued for predicates in $S$.
\end{myproof}}

\noindent The inclusion in Theorem \ref{thm:comparison-supported-subset} is strict for the program $\pgm_3$ described above.  $\constraint(\pgm_3) = \{ \{\NOT\,\co{q}\} \}$ and $\supported(\pgm_6) = \{ \{\co{q}\}, \{\NOT\,\co{q}\} \}$. 
 
\myparag{Stable models}
Gelfond and Lifschitz define {\em stable model semantics} (SMS) of logic programs \cite{GelLif88}. They define the {\em stable models} of a program $\pgm$ to be the 2-valued interpretations of $\pgm$ that are fixed points of a particular transformation.\full{  Van Gelder et al.\ proved that the stable models of $\pgm$ are exactly the 2-valued fixed points of the operator $W_\pgm$ described above \cite[Theorem 5.4]{van+91well}.}  Let $\stable(\pgm)$ denote the set of stable models of $\pgm$. 

\begin{mytheorem}
  For every program $\pgm$, $\stable(\pgm) \subseteq \constraint(\pgm)$.
\label{thm:comparison-stable}
\end{mytheorem}

\newcommand{\thmComparisonStableProof}{
\noindent
\begin{myproof}{ of Theorem \ref{thm:comparison-stable}}
  Let $M\in\stable(\pgm)$.  We need to show  $M\in\constraint(\pgm)$, i.e., $\founded(\pgm) \subseteq M$ and $M$ is a 2-valued model of $\cmpl(\pgm)$.  By Theorem \ref{thm:comparison-wf}, $\founded(\pgm)\subseteq\wf(\pgm)$.  $M$ is a fixed point of $W_\pgm$ \cite[Theorem 5.4]{van+91well}, so $\wf(\pgm) \subseteq M$.  By transitivity, $\founded(\pgm)\subseteq M$.  It is easy to show that $M$ is a 2-valued model of $\cmpl(\pgm)$ iff it is a 2-valued model of $\pgm$ and $\cmpl_N(\pgm)$, where $\cmpl_N(\pgm)$ denotes the completion rules added by $\addinv$ (``N'' reflects the negative conclusions).  Gelfond and Lifschitz proved that every stable model of $\pgm$ is a 2-valued model of $\pgm$ \cite[Theorem 1]{GelLif88}.  It remains to show that $M$ is a model of $\cmpl_N(\pgm)$.  
  Let $\neg P~\IF~\neg H_1 \land \cdots \land \neg H_n$ be a rule in $\cmpl_N(\pgm)$.  We need to show that, if $M$ satisfies $\neg H_1 \land \cdots \land \neg H_n$, then $M$ satisfies $\neg P$.  It suffices to show $P \in U_\pgm(M)$, because this implies $\neg P \in M$.  The rules defining $P$ in $\pgm$ have the form $P~\IF~H_i$, for $i\in [1..n]$, and each $H_i$ is false in $M$ by assumption, so some conjunct in each $H_i$ is false in $M$, so by definition of unfounded set, $P \in U_\pgm(M)$.
\end{myproof}}

\noindent The inclusion in Theorem \ref{thm:comparison-stable} is strict for program 7 in Table \ref{tab-sem}, denoted $\pgm_7$, which has two rules \co{q \IF $\NOT\,$q} and \co{q \IF q}. $\stable(\pgm_7) = \emptyset$ and $\constraint(\pgm_7) = \{\co{q}\}$.


\newcommand{\selfFalseSemantics}{
\myparag{Closed predicate assumption}
We can extend the language to support declaration of \uncert complete
predicates as {\it closed}.  Informally, this means that an atom $A$ of the
predicate is false in an interpretation $I$, called {\it self-false} in $I$, if every ground instance of rules that concludes $A$, or recursively concludes some hypothesis of that rule instance, has a hypothesis that is false or, recursively, is self-false in $I$.  Self-false atoms are elements of unfounded sets\myconf{ \cite{van+91well}}.


Formally, $\selffalse_\pgm(I)$, the set of self-false atoms of program $\pgm$ with respect to interpretation $I$, is defined in the same way as the greatest unfounded set of $\pgm$ with respect to $I$, except replacing
``some positive hypothesis of $R$ is in $\ufs$'' with ``some positive hypothesis of $R$ for a closed predicate is in $\ufs$''.  The founded semantics of this extended language is defined by repeatedly computing the semantics as per Section \ref{sec-sem} and then setting self-false atoms to false, until a least fixed point is reached.  Formally, the founded semantics is $\foundedcl(\pgm)=\lfp(F_\pgm)$, where $F_\pgm(I) = \founded(\pgm \union I) \union \neg \cdot \selffalse_\pgm(\founded(\pgm \union I))$.

The constraint semantics for this extended language includes only interpretations that contain the negative literals
required by the closed declarations.  Formally, a {\em constraint model} of a program
$\pgm$ with closed declarations is a consistent 2-valued interpretation
$M$ such that $M$ is a model of
$\cmpl(\pgm)$, $\foundedcl(\pgm) \subseteq M$, and $\neg \cdot \selffalse_\pgm(M) \subseteq M$.  Let
$\constraintcl(\pgm)$ denote the set of constraint models of $\pgm$.


The next theorem states that changing predicate declarations from \uncert, complete, and closed to \cert when allowed, or vice versa, preserves founded and constraint semantics.  Theorem \ref{thm:scc-certainty} implies that this change needs to be made for  all predicates in an SCC.

\begin{mytheorem}
Let $\pgm$ be a program.  Let $S$ be an SCC in its dependence graph containing only predicates that are \uncert, complete, and closed.  Let $\pgm'$ be a program identical to $\pgm$ except that all predicates in $S$ are declared certain.  Note that, for the declarations in both programs to be allowed, predicates in all SCCs that follow $S$ in dependency order must be uncertain, predicates in all SCCs that precede $S$ in dependency order must be \cert, and predicates in $S$ must not have circular negative dependency.
Then $\foundedcl(\pgm)=\foundedcl(\pgm')$ and $\constraintcl(\pgm)=\constraintcl(\pgm')$.
\label{thm:uncertain-certain}
\end{mytheorem}
\newcommand{\thmUncertainCertainProof}{
\noindent
  \begin{myproof}{ of Theorem \ref{thm:uncertain-certain}}
    First, we show that $\foundedcl(\pgm)$ is 2-valued for predicates in $S$.  Let $RS$ be the set of all instances of combined rules and completion rules in $\cmpl(\pgm)$ for predicates in $S$.  Note that every positive literal and negative literal for every predicate in $S$ appears as the conclusion of at least one rule in $G$ (this holds even if rules in $\pgm$ contain conclusions of the form \co{p(x,x)} or \co{p(x,0)}, due to the fresh variables and existential quantifiers introduced by $\comb$).  Let $\UA$ be the set of atoms for predicates in $S$ that are undefined in $\founded(\pgm)$.  For each atom $A$ in $\UA$, since the predicate in $A$ is complete and $A$ is not \T or \F in $\founded(\pgm)$, (1) for every rule $R$ in $RS$ with conclusion $A$, some hypothesis of $R$ is \F or \UD in $\founded(\pgm)$, and (2) for some rule $R$ in $RS$ with conclusion $A$, some hypothesis of $R$ is \UD in $\founded(\pgm)$.  Since all predicates in SCCs that precede $S$ are certain, these undefined hypotheses must be atoms in $\UA$ or their negations.  Define a dependence relation $\rightarrow$ on $\UA$ by: $B \rightarrow A$ if some rule $R$ in $RS$ with conclusion $A$ has an undefined hypothesis that is $B$ or $\neg B$.  The previous observation implies that, for every $A$ in $\UA$, there exists $B$ in $\UA$ such that $B \rightarrow A$.  Since $\UA$ is finite, this implies that every atom in $\UA$ is in a $\rightarrow$-cycle.  Since predicates in $S$ do not have circular negative dependency, this implies that all hypotheses involved in the cycle are positive.  These observations, together with all predicates in $S$ being closed, imply that the literals in every cycle, and hence all atoms in $\UA$, are in $\selffalse_\pgm(\founded(\pgm))$.  This implies that $\foundedcl(\pgm)$ contains the negations of all literals in $\UA$.  Therefore, $\foundedcl(\pgm)$ is 2-valued.

    Next, we show that $\foundedcl(\pgm)=\foundedcl(\pgm')$.  For each predicate $P$ in $S$, the two programs contain equivalent rules for adding positive literals for $P$ to the founded model, because the combined rule for $P$ in $\pgm$ is logically equivalent to the original rules for $P$ in $\pgm$, so $\foundedcl(\pgm)$ and $\foundedcl(\pgm')$ contain the same positive literals for $P$.  Since both models are 2-valued for $P$, they also contain the same negative literals for $P$.

    Finally, we show that $\constraintcl(\pgm)=\constraintcl(\pgm')$.  We consider the three conditions in the definition of $\constraintcl$, in turn.

 Consider the first condition, namely, $\foundedcl(\pgm)\subseteq M$.  It is equivalent for the two programs, because they have the same founded model.

 Consider the second condition, namely, $M$ satisfies $\cmpl(\pgm)$.  It differs for the two programs in that $\cmpl(\pgm)$ contains combined rules and completion rules for predicates in $S$, while $\cmpl(\pgm')$ contains the original rules in $\pgm$ for predicates in $S$.  Since $\foundedcl(\pgm)=\foundedcl(\pgm')$, Theorem \ref{thm:consistent} implies $\foundedcl(\pgm)$ is a model of $\cmpl(\pgm)$ and $\cmpl(\pgm')$.  Since $\foundedcl(\pgm)$ is 2-valued for predicates in $S$ and all predicates on which they depend, it is 2-valued for all predicates used in those rules.  Therefore, every $M$ satisfying $\foundedcl(\pgm)\subseteq M$ contains the same literals as $\foundedcl(\pgm)$ for all predicates used in those rules, so $M$ satisfies $\cmpl(\pgm)$ and $\cmpl(\pgm')$. Thus, the second condition is equivalent for the two programs.

 Consider the third condition, namely, $\neg \cdot \selffalse_\pgm(M) \subseteq M$.  $\foundedcl(\pgm)$ is 2-valued for predicates in $S$ and all predicates on which they depend, so for every $M$ satisfying $\foundedcl(\pgm)\subseteq M$, $\foundedcl(\pgm)$ and $M$ contain the same literals for those predicates, so $\selffalse_{\pgm}(M) = \selffalse_\pgm(\foundedcl(\pgm))$ and $\selffalse_{\pgm'}(M) = \selffalse_{\pgm'}(\foundedcl(\pgm))$.  
 For every instance $R$ of a rule whose conclusion is for a predicate in $S$, every hypothesis of $R$ is \T or \F (not undefined) in $\foundedcl(M)$, so disjunct (2) in the definition of self-false set cannot be used to treat any additional hypothesis of $R$ as \F, regardless of which predicates are closed, so $\selffalse_\pgm(\foundedcl(\pgm))=\selffalse_{\pgm'}(\foundedcl(\pgm))$.  Using these equalities and transitivity, $\selffalse_\pgm(M) = \selffalse_{\pgm'}(M)$.  Thus, the third condition is equivalent for the two programs.
  \end{myproof}
}


 \begin{mytheorem}
  For a program $\pgm$ in which every predicate is uncertain, complete, and closed, $\foundedcl(\pgm)=\wf(\pgm)$.
   \label{thm:self-false-WFS}
 \end{mytheorem}
   
 \begin{mytheorem}
   For a program $\pgm$ in which every predicate is uncertain, complete, and closed, $\constraintcl(\pgm) = \stable(\pgm)$.
 \label{thm:self-false-SMS}
 \end{mytheorem}
}

\newcommand{\thmSelfFalseWFSProof}{
\noindent
\begin{myproof}{ of Theorem \ref{thm:self-false-WFS}}
  First, we show $\foundedcl(\pgm)\subseteq\wf(\pgm)$, by proving by induction on the computation of the least fixed point that, in each step, $F_\pgm(I) \subseteq \wf(\pgm)$.  
The induction hypothesis is $I \subseteq \wf(\pgm)$, and we need to show $F_\pgm(I) \subseteq \wf(\pgm)$.  It suffices to show (1) $\founded(\pgm \union I) \subseteq \wf(\pgm)$ and (2) $\neg \cdot \selffalse_\pgm(\founded(\pgm \union I)) \subseteq \wf(\pgm)$, since $F_\pgm(I)$ is the union of these two sets.

Proof of (1): By Theorem \ref{thm:comparison-wf}, $\founded(\pgm \union I) \subseteq \wf(\pgm \union I)$.  It is easy to show that, for any subset $I$ of $\wf(\pgm)$, $\wf(\pgm\union I) = \wf(\pgm)$.  Thus, $\founded(\pgm \union I) \subseteq \wf(\pgm)$.

Proof of (2): $\selffalse_\pgm$ is monotone, and $\founded(\pgm \union I) \subseteq \wf(\pgm)$ from (1),\linebreak[4] so $\selffalse_\pgm(\founded(\pgm \union I)) \subseteq \selffalse_\pgm(\wf(\pgm))$.  $\selffalse_\pgm$ is $U_\pgm$ restricted to specified predicates, so $\selffalse_\pgm(I) \subseteq U_\pgm(I)$ for any interpretation $I$.  Thus,  $\selffalse(\founded(\pgm \union I)) \subseteq U_\pgm(\wf(\pgm))$.  By definition of $\wf_\pgm$, $\neg \cdot U_\pgm(\wf(\pgm)) \subseteq \wf(\pgm)$.  Thus, $\neg \cdot \selffalse(\founded(\pgm \union I)) \subseteq \wf(\pgm)$.

Second, we show $\wf(\pgm)\subseteq\foundedcl(\pgm)$, by proving by induction on the computation of the least fixed point that, in each step, $W_\pgm(I) \subseteq \foundedcl(\pgm)$.  The induction hypothesis is $I \subseteq \foundedcl(\pgm)$, and we need to show $W_\pgm(I) \subseteq \foundedcl(\pgm)$.  It suffices to show (a) $T_\pgm(I) \subseteq \foundedcl(\pgm)$ and (b) $\neg \cdot U_\pgm(I) \subseteq \foundedcl(\pgm)$, since $W_\pgm(I)$ is the union of these two sets.  

The proof of (a) is straightforward using the induction hypothesis and the definitions of $T_\pgm$ and $\founded$. 
The proof of (b) uses the following lemma relating unfounded sets and self-false sets, which is easily proved from the definitions: $U_\pgm(I) \subseteq \selffalse_\pgm(I)$ when all predicates are uncertain, complete, and closed.  
The induction hypothesis for (b) is $I \subseteq \foundedcl(\pgm)$.  $U_\pgm$ is monotone, so $U_\pgm(I) \subseteq U_\pgm(\foundedcl(\pgm))$.  The above lemma implies $U_\pgm(\foundedcl(\pgm)) \subseteq \selffalse_\pgm(\foundedcl(\pgm))$.  
By transitivity, $U_\pgm(I) \subseteq \selffalse_\pgm(\foundedcl(\pgm))$.  By definition of $\foundedcl$, $\neg\cdot \selffalse_\pgm(\foundedcl(\pgm))
\subseteq \foundedcl(\pgm)$.  Using this inequality, the preceding
inequality, and transitivity, we conclude $\neg\cdot U_\pgm(I) \subseteq
\foundedcl(\pgm)$.
\end{myproof}}

\newcommand{\thmSelfFalseSMSProof}{
\noindent
\begin{myproof}{ of Theorem \ref{thm:self-false-SMS}}
  Proof that $\stable(\pgm) \subseteq \constraintcl(\pgm)$.  The proof uses Theorem \ref{thm:comparison-stable}, which has the additional hypothesis that all predicates have default declarations as \cert or \uncert.  Theorem \ref{thm:comparison-stable} is applicable here nevertheless, because that additional hypothesis is unnecessary in the context of the other hypotheses of this theorem.  To see this, note that non-default declarations of predicates as \cert or \uncert can differ from the default declarations only by unnecessarily declaring some predicates \uncert.  By hypothesis, those predicates are also declared complete and closed.  By applying Theorem \ref{thm:uncertain-certain} to each SCC containing those predicates, we conclude that these non-default declarations do not change the founded and constraint semantics.  Therefore, we can assume in the remainder of the proof that predicates have their default declarations as certain or uncertain.
 
  Let $M \in \stable(\pgm)$.  We need to show $M \in \constraintcl(\pgm)$, i.e., (1) $\foundedcl(\pgm) \subseteq M$, (2) $M$ is a model of $\cmpl(\pgm)$, and (3) $\neg \cdot \selffalse_\pgm(M) \subseteq M$.
 
Proof of (1): As shown in the proof of Theorem \ref{thm:comparison-stable}, $\wf(\pgm)\subseteq M$.  By Theorem \ref{thm:self-false-WFS},\nottlp{\linebreak} $\foundedcl(\pgm) = \wf(\pgm)$, so $\foundedcl(\pgm) \subseteq M$.  Proof of (2): Same as in the proof of Theorem \ref{thm:comparison-stable}.  Proof of (3): Every \uncert predicate is closed, so $\selffalse_\pgm(M) \subseteq U_\pgm(M)$, so $\neg \cdot \selffalse_\pgm(M) \subseteq W_\pgm(M)$.  $M$ is a fixed point of $W_\pgm$ \cite[Theorem 5.4]{van+91well}, so $W_\pgm(M)=M$.  Using this to simplify the right side of the previous inequality, we conclude $\neg \cdot \selffalse_\pgm(M) \subseteq M$.

Proof that $\constraintcl(\pgm) \subseteq \stable(\pgm)$.  Let $M \in \constraintcl(\pgm)$.  We need to show $M \in \stable(\pgm)$; this is equivalent to showing $M$ is a fixed point of $W_\pgm$ \cite[Theorem 5.4]{van+91well}.  We prove $W_\pgm(M) \subseteq M$ and $M \subseteq W_\pgm(M)$.

Proof that $W_\pgm(M) \subseteq M$: We need to show $T_\pgm(M)\subseteq M$ and $\neg \cdot U_\pgm(M) \subseteq M$.  The former follows from the fact that $M$ is a model of $\cmpl(\pgm)$.  

The latter follows from $\neg \cdot \selffalse_\pgm(M) \subseteq M$ and $\selffalse_\pgm(M) = U_\pgm(M)$, as shown next.  Since the definition of $\selffalse_\pgm(M)$ is obtained from the definition of $U_\pgm(M)$ by limiting in the recursive disjunct to closed predicates, $\selffalse_\pgm(M) \subseteq U_\pgm(M)$ always holds.  Since all \uncert predicates are also closed, to show $\selffalse_\pgm(M) = U_\pgm(M)$, it suffices to show that, for each atom $A$ in $U_\pgm(M)$ for a \cert predicate $P$, $A \in \selffalse_\pgm(M)$.  To see this, note that $\neg A \in \foundedcl_\pgm(M)$, because $\foundedcl(M)$ includes all negative literals for \cert predicates that can be in any consistent model of $\pgm$ (recall that \cert predicates cannot depend on \uncert predicates, so this holds regardless of undefined values in $\foundedcl(M)$).  Since $\neg A \in \foundedcl_\pgm(M)$ and $\foundedcl_\pgm(M)\subseteq M$, we have $\neg A \in M$ and hence $A \in \selffalse_\pgm(M)$.  

Proof that $M \subseteq W_\pgm(M)$: Consider any literal in $M$.  We need to show that the literal is in $W_\pgm(M)$.

Case 1: Consider a positive literal $A$ in $M$.  We show $A \in T_\pgm(M)$ hence $A \in W_\pgm(M)$.

Case 1.1: $A$ is for a \cert predicate.  $\foundedcl(M)$ and $M$ contain the same literals for such predicates, so $A \in \foundedcl(M)$, so $A \in T_\pgm(I)$, where $I$ is the intermediate approximation to $\foundedcl(M)$ at the step when $A$ is added.  $T_\pgm$ is monotonic, and $I \subseteq \foundedcl(M) \subseteq M$, so $A \in T_\pgm(M)$.

Case 1.2: $A$ is for a \uncert predicate $P$.  $M$ satisfies $\cmpl(\pgm)$.  $\cmpl(M)$ contains the combined rule $C~\IF~B$ for $P$ and its inverse $\neg C~\IF~\neg B$.   $M$ is 2-valued, and in 2-valued models, the conjunction of these two rules is equivalent to $C~\Leftrightarrow~B$.  Therefore, $A$ is derivable in $M$ using an instance of the combined rule for $P$, which is logically equivalent to the original rules for $P$ in $\pgm$, so $A \in T_\pgm(M)$.

Case 2: consider a negative literal $\neg A$ in $M$.  We show $A \in U_\pgm(M)$ hence $\neg A \in W_\pgm(M)$.

Case 2.1: $\neg A$ is for a \cert predicate.  $\foundedcl(M)$ and $M$ contain the same literals for such predicates, so $\neg A \in \foundedcl(\pgm)$.  By Theorem \ref{thm:self-false-WFS}, $\foundedcl(\pgm)=\wf(\pgm)$, so $\neg A \in \wf(\pgm)$, so $A \in U_\pgm(\foundedcl(\pgm))$, so by monotonicity, $A \in U_\pgm(M)$.

Case 2.2: $\neg A$ is for a \uncert predicate $P$.  By reasoning similar to case 1.2, $\neg A$ is derivable in $M$ using an instance of the inverse of the combined rule for $P$.  By definition of the combined rule, this implies that, for every instance with conclusion $A$ of a rule for $P$ in $\pgm$, the body of the rule evaluates to false in $M$.  This implies $A \in U_\pgm(M)$.
\end{myproof}}


%% file: smallexamp.tex
\mysec{Comparison of semantics for well-known small examples and more}
\label{sec-small}

Table~\ref{tab-sem} shows well-known example rules and more for tricky
boundary cases in the semantics, where all \uncert predicates that are
in a conclusion are declared complete, but not closed, and shows
different semantics for them.
\notlfcs{\begin{table*}[htb]}
\lfcs{\begin{sidewaystable}[htb]}
\lfcs{
}

  \small
  \centering

  \begin{\nottlp{tabular}\tlp{oldtabular}}{@{\,}c@{~}|@{~}c@{~}||c@{~}|@{~}c@{~}|c@{~}|@{~}c@{~}||c@{~}|@{~}c@{~}|c@{~}|@{~}c@{}}
    \mbox{~}
    & Program
    & \multicolumn{2}{c|}{Founded}     & WFS   & Fitting
    & \multicolumn{2}{c|}{Constraint}  & SMS   & Supported\\

    \mbox{~}
    &
    & \multicolumn{2}{c|}{(not closed)} & & (Kripke  
    & \multicolumn{2}{c|}{(not closed)} & &  \\

    \cline{3-4}     \cline{7-8}

    \mbox{~}
    &
    & \uncert      & \cert        & & -Kleene)
    & \uncert      & \cert        & & \\
    \hline\hline

    1
    & q \IF \NOT q 
    & \{\uk q\}    & \NA          & \{\uk q\}   & \{\uk q\}   
    & no model     & \NA          & no model    & no model\\
    \hline

    2
    &\begin{\nottlp{tabular}\tlp{oldtabular}}[c]{@{}l@{}}
      q \IF \NOT p\\
      p \IF \NOT q
    \end{\nottlp{tabular}\tlp{oldtabular}}    
    &\{\uk p, \uk q\} & \NA        &\{\uk p, \uk q\}& \{\uk p, \uk q\}
    & \{p, \fa q\},\{\fa p, q\} & \NA & \{p, \fa q\},\{\fa p, q\} & \{p, \fa q\},\{\fa p, q\}\\
    \hline\hline

    3
    & q \IF q \hfill\mbox{~}
    & \{\uk q\}    & \{\fa q\}    & \{\fa q\}   & \{\uk q\}
    & \{q\},\{\fa q\} & \{\fa q\} & \{\fa q\}   & \{q\},\{\fa q\}\\
    \hline

    4
    &\begin{\nottlp{tabular}\tlp{oldtabular}}[c]{@{}l@{}}
      q \IF p\\
      p \IF q
    \end{\nottlp{tabular}\tlp{oldtabular}} \hfill\mbox{~}
    &\{\uk p, \uk q\} & \{\fa p, \fa q\} & \{\fa p, \fa q\} & \{\uk p, \uk q\}
    & \{p, q\},\{\fa p, \fa q\} & \{\fa p, \fa q\} & \{\fa p, \fa q\} & \{p, q\},\{\fa p, \fa q\}\\
    \hline\hline

    5
    & q \IF \NOT p
    &\{\uk p, \uk q\} & \{\fa p, q\} & \{\fa p, q\} & \{\fa p, q\}
    & \{p, \fa q\},\{\fa p, q\}  & \{\fa p, q\} & \{\fa p, q\}       & \{\fa p, q\}\\
    \hline

    6 
    & q \IF p \hfill\mbox{~}
    & \{\uk p, \uk q\} & \{\fa p, \fa q\} &  \{\fa p, \fa q\} & \{\fa p, \fa q\}
    & \{p, q\},\{\fa p, \fa q\} & \{\fa p, \fa q\} & \{\fa p, \fa q\} & \{\fa p, \fa q\}\\
    \hline\hline

    7
    &\begin{\nottlp{tabular}\tlp{oldtabular}}[c]{@{}l@{}}
      q \IF \NOT q\\
      q \IF q
    \end{\nottlp{tabular}\tlp{oldtabular}}    
    & \{\uk q\}    & \NA          & \{\uk q\}   & \{\uk q\}
    & \{q\}        & \NA          & no model    & \{q\}\\
    \hline

    8
    &\begin{\nottlp{tabular}\tlp{oldtabular}}[c]{@{}l@{}}
      q \IF \NOT q\\
      \mbox{~}\hfill \AND q
    \end{\nottlp{tabular}\tlp{oldtabular}}    
    & \{\uk q\}    & \NA          & \{\fa q\}   & \{\uk q\}
    & \{\fa q\}    & \NA          & \{\fa q\}   & \{\fa q\}\\
    \hline

  \end{\nottlp{tabular}\tlp{oldtabular}}

  \caption{Different semantics for programs where all \uncert predicates 
    that are in a conclusion are declared complete, but not closed.
    ``\uncert'' means all predicates in the program are declared \uncert.
    ``\cert'' means all predicates in the program that can be
    declared \cert are declared \cert;
    ``\NA'' means no predicates can be declared \cert, 
    so the semantics is the same as ``\uncert''.  
    p, \fa p and \uk p mean p is $\T$, $\F$, and $\UD$, respectively.}

\lfcs{
}
  \label{tab-sem}

\lfcs{\end{sidewaystable}}
\notlfcs{\end{table*}}

\begin{itemize}

\item Programs 1 and 2 contain only negative cycles.  All three of
  Founded, WFS, and Fitting agree.  All three of Constraint, SMS, and
  Supported agree.

\item Programs 3 and 4 contain only positive cycles.  Founded for
  \cert agrees with WFS; Founded for \uncert agrees with Fitting.
  Constraint for \cert agrees with SMS; Constraint for \uncert agrees
  with Supported.

\item Programs 5 and 6 contain no cycles.  Founded for \cert agrees
  with WFS and Fitting; Founded for \uncert has more undefined.
  Constraint for \cert agrees with SMS and Supported; Constraint for
  \uncert has more models.

\item Programs 7 and 8 contain both negative and positive cycles.  For
  program 7 where \co{\NOT q} and \co{q} are disjunctive, all three of
  Founded, WFS, and Fitting agree; Constraint and Supported agree, but SMS
  has no model.
  For program 8 where \co{\NOT q} and \co{q} are conjunctive, Founded and
  Fitting agree, but WFS has \co{q} being \F; all three of Constraint, SMS,
  and Supported agree.

\end{itemize}
For all 8 programs, with default complete but not closed predicates, we
have the following:
\begin{itemize}

\item If all predicates are the default \cert or \uncert, then Founded
  agrees with WFS, and Constraint agrees with SMS, with one exception
  for each:

  (1) Program 7 concludes \co{q} whether \co{q} is \F or \T, so SMS
  having no model is an extreme outlier among all 6 semantics and is
  not consistent with common sense.

  (2) Program 8 concludes \co{q} if \co{q} is \F and \T, so Founded
  semantics with \co{q} being \UD is imprecise, but Constraint has \co{q}
  being \F.  WFS has \co{q} being \F because it uses \F for ignorance.

\item If predicates not in any conclusion are \cert (not shown in
  Table~\ref{tab-sem} but only needed for \co{p} in programs 5 and
  6), and other predicates are \uncert, then Founded equals Fitting,
  and Constraint equals Supported, as captured in
  Theorems~\ref{thm:comparison-fitting-equal} and
  \ref{thm:comparison-supported-equal}, respectively.

\item If all predicates are \uncert, then Founded has all values being
  \UD, capturing the well-known unclear situations in all these
  programs, and Constraint gives all different models except for
  programs 2 and 5, and programs 4 and 6, which are pair-wise
  equivalent under completion, capturing exactly the differences among
  all these programs.

\end{itemize}
Finally, if all predicates in these programs are not complete, then Founded and Constraint are the same as in Table~\ref{tab-sem} except that Constraint for \uncert becomes equivalent to truth values in first-order logic: 
programs 1 and 8 have an additional model, {\{q\}, program 6 has an additional model, \{\fa p, q\}, and programs 2 and 5 have an additional model, \{p,q\}.

%% file: addexamp.tex
\mysec{Additional examples}
\label{sec-examp}

We discuss the semantics of several main well-known examples.

\myparag{Win-not-win game}
This is the running example with the \co{move} facts and a single rule that defines \co{win} recursively using its own negation.

With the default declaration that \co{move} is certain and \co{win} is uncertain and complete, founded semantics and constraint semantics are as discussed in Section~\ref{sec-sem}.
There is no circular positive dependency.
Fitting semantics and WFS are the same as founded semantics, and supported model semantics and SMS are the same as constraint semantics.

If \co{move} is declared \uncert instead of the default of being \cert, then moves not in the given \co{move} facts have \UD values, not allowing any \co{n.win} or \co{win} facts to be inferred.   Therefore, founded semantics infers \co{move} to be \T for the given \co{move} facts and \UD for all other pairs of positions, and \co{win} to be \UD for all positions.
Constraint semantics extends the \T values for \co{move} with all combinations of \T and \F values for the \UD values of \co{move} and \co{win} such that the combination satisfies the given rule and completion rule.

\myparag{Graph reachability}
A source vertex \co{x} is represented by a fact \co{source(x)}.  An
edge from a vertex \co{x} to a vertex \co{y} is represented by a fact
\co{edge(x,y)}.  The following two rules capture graph reachability,
i.e., the set of vertices reachable from source vertices by following
edges.
\begin{code}
    reach(x) \IF source(x)
    reach(y) \IF edge(x,y) \AND reach(x)
\end{code}

In the dependency graph, each predicate is in a separate SCC, and the SCC for \co{reach} is ordered after the other two.  There is no negation in this program.

With the default declarations of all predicates being \cert, no
completion rules are added.  The least fixed point computation for
founded semantics infers \co{reach} to be \T for all vertices that are
source vertices or are reachable from source vertices by following
edges, as desired.  For the remaining vertices, \co{reach} is \F.
Constraint semantics is the same.
These are the same as WFS and SMS.

If \co{reach} is declared \uncert and complete, but not closed, then
after completion, we obtain
\begin{code}
    reach(x) \IF source(x) \OR \myconf{
             }(\SOME y | (edge(y,x) \AND reach(y)))
    n.reach(x) \IF n.source(x) \AND \myconf{
             }(\EACH y | (n.edge(y,x) \OR n.reach(y)))
\end{code}
The least fixed point computation for founded semantics infers
\co{reach} to be \T for all reachable vertices as when predicates are
\cert, and infers \co{reach} to be \F for all vertices that are not
source vertices and that have no in-coming edge at all or have
in-coming edges only from vertices for which \co{reach} is \F.  For
the remaining vertices, i.e., those that are not reachable from the
source vertices but are in cycles of edges, \co{reach} is \UD.  Constraint semantics extends this model with all combinations of \T and \F values for the \UD values such that the combination satisfies the given rules and completion rule.
These are
the same as in Fitting semantics and supported model semantics, respectively.

\myparag{Russell's paradox}
Russell's paradox is well known as the barber paradox.  The barber is a
man who shaves all those men, and those men only, who do not shave
themselves, specified using the following fact and rule:  %
\begin{code}
    man('barber')
    shave('barber',x) \IF man(x) \AND \NOT shave(x,x)
\end{code}
The question is: Does the barber shave himself?  That is: What is the value
of
\co{shave('barber',} \co{'barber')}?

With the default declarations that \co{man} is \cert, and \co{shave} is
\uncert (because \co{shave} is defined recursively using its
own negation) and complete, which captures "those men only",
the completion step adds the rule
\begin{code}
    \NOT shave(y,x) \IF y\,\NEQ{\,}'barbar' \OR \NOT man(x) \OR shave(x,x)
\end{code}
The completed program, after eliminating negation, is
\begin{code}
    shave('barber',x) \IF man(x) \AND n.shave(x,x)
    man('barber')
    n.shave(y,x) \IF y\,\NEQ{\,}'barbar' \OR n.man(x) \OR shave(x,x))
\end{code}
The least fixed point computation for founded semantics infers no \T
or \F facts of \co{shave}, leaving \co{shave('barber','barber')} to be \UD.
Constraint semantics has no model.  These results correspond to WFS
and SMS, respectively.  %
Our results show the exact assumptions (certain and complete) and desired outcome restrictions (allowing \UD values or allowing only \T and \F values) that lead to different solutions. 

If there are other men besides the barber, then founded semantics 
also infers %
\co{shave('barber',x)} for all man \co{x} except \co{'barber'} to be \T, and %
\co{shave(y,x)} for all man \co{y} except \co{'barber'} and for all man
\co{x} to be \F, %
leaving only \co{shave('barber','barber')} to be \UD.
For example, if there is also a fact \co{man('tom')}, then founded semantics also infers \co{shave('barber','tom')} to be \T, and \co{shave('tom','barber')} and \co{shave('tom','tom')} to be \F.
Constraint semantics has no model.
These results again correspond to WFS and SMS, respectively,
and show the exact assumptions and outcome restrictions that lead to different solutions.

\myparag{Even numbers}
In this example, even numbers are defined by the predicate \co{even}, and natural numbers in order
are given using the predicate \co{succ}.
\begin{code}
    even(n) \IF succ(m,n) \AND \NOT even(m)
    even(0)
    succ(0,1)
    succ(1,2)
    succ(2,3)
\end{code}

With the default declarations that \co{succ} is \cert and \co{even} is
\uncert and complete, the following completion rule is added:
\begin{code}
    n.even(n) \IF n\,\NEQ{\,}0 \AND (\EACH m | n.succ(m,n) \OR even(m))
\end{code}
Founded semantics infers that \co{even(1)} is
\F, \co{even(2)} is \T, and \co{even(3)} is \F.  Constraint semantics
is the same.  These results are the same as WFS and SMS.

\myparag{Yale shooting}
This example is about whether a turkey is alive, given some facts
and rules about whether and when a gun is loaded, specified below.  It
uses the extension that allows negative facts and negative
conclusions.
\begin{code}
    alive(0)
    \NOT loaded(0)
    loaded(1)
    \NOT alive(3) \IF loaded(2)
\end{code}

Assume both predicates \co{alive} and \co{loaded} are declared \uncert 
and not complete.  
In the dependency graph, there are two SCCs: one with \co{loaded}, one
with \co{alive}, and the former is ordered before the latter.
Founded semantics infers that
\co{loaded(0)} is \F, \co{loaded(1)} is \T, \co{loaded(2)} and
\co{loaded(3)} are \UD, \co{alive(0)} is \T, and \co{alive(1)},
\co{alive(2)}, and \co{alive(3)} are \UD.
Constraint semantics has multiple models,
some containing that \co{loaded(2)} is \T and \co{alive(3)} is \F,
and some containing that \co{loaded(2)} is \F and \co{alive(3)} is \T.
These confirm the well-known outcomes.

If there are other facts and rules that give or infer \co{loaded(2)} to be \T, then \co{alive(3)} is \F in both founded and constraint semantics.  This is again the well-known outcome.

\myparag{Variant of Yale shooting}
This is a variant of the Yale shooting problem, copied from~\cite{van+91well}:%
\begin{code}
    noise(T) \IF loaded(T) \AND shoots(T).
    loaded(0).
    loaded(T) \IF succ(S,T) \AND loaded(S) \AND \NOT shoots(S).
    shoots(T) \IF triggers(T).
    triggers(1).
    succ(0,1).
\end{code}

There is no circular negative dependency, so the default is that all predicates are \cert.  In this case, no completion rules are added.
Founded semantics and constraint semantics both yield that
\co{succ(0,1)} and
\co{loaded(0)} are \T;
\co{trigger(0)},
\co{shoots(0)}, and
\co{noise(0)} are \F; and
\co{loaded(1)}, 
\co{trigger(1)}, 
\co{shoots(1)}, and
\co{noise(1)} are \T. 
This is the same as WFS, Fitting semantics, SMS, and supported models.

%% file: appendix-proofs.tex
\proofs{
\section{Proofs}
\label{sec-proofs}

\thmConsistentProof
\medskip

\thmModelProof
\medskip

\thmEquivDeclsProof
\medskip

\thmSCCcertaintyProof
\medskip

\thmPredMergeProof
\medskip

\thmComparisonStratifiedProof
\medskip

\thmComparisonLogicProof
\medskip

\thmComparisonFittingEqualProof
\medskip

\thmComparisonFittingSubsetProof
\medskip

\thmComparisonFittingSupsetProof
\medskip

\thmComparisonWfProof
\medskip

\thmComparisonSupportedEqualProof
\medskip

\thmComparisonSupportedSupsetProof
\medskip

\thmComparisonSupportedSubsetProof
\medskip

\thmComparisonStableProof
\medskip

\thmUncertainCertainProof
\medskip

\thmSelfFalseWFSProof
\medskip

\thmSelfFalseSMSProof

}
